\documentclass[aps,prl,twocolumn]{revtex4}

\begin{document}


\title{Cosmological expansion and local physics}


\author{Valerio Faraoni}
\email[]{vfaraoni@ubishops.ca}
\author{Audrey Jacques}
\email[]{ajacques@ubishops.ca}
\affiliation{Physics Department, Bishop's University\\
2600 College Street, Sherbrooke, Qu\'{e}bec, Canada J1M~0C8
}


\date{\today}

\begin{abstract}
The interplay between cosmological expansion and local 
attraction 
in a gravitationally bound system is revisited in various 
regimes. First, weakly gravitating Newtonian systems are 
considered, followed by various exact solutions describing a 
relativistic central object embedded in a Friedmann universe. It 
is shown that the ``all or nothing'' behaviour  recently 
discovered (i.e., weakly coupled systems are comoving while 
strongly coupled ones resist the cosmic expansion) is limited 
to the de Sitter background. New exact solutions are presented 
which describe black holes perfectly comoving with a generic 
Friedmann universe. The possibility of violating cosmic 
censorship for a black hole approaching the Big Rip is also 
discussed.
\end{abstract}


\maketitle

\section{Introduction}
\setcounter{equation}{0}

The issue of whether a planet, a star, or a  galaxy expands 
following the rest of the universe is a problem of principle in 
general relativity  that still awaits a definitive answer. The 
effect of the cosmological expansion on local systems such as 
the Solar System has  a long history dating back to the 1933 
paper by McVittie \cite{McVittie} introducing  a spacetime 
metric that 
represents a point mass embedded in a 
Friedmann-Lemaitre-Robertson-Walker (FLRW) universe. Later work 
by Einstein and Straus \cite{EinsteinStraus} introduced the 
Swiss-cheese model which is, however, unable to describe the 
Solar System \cite{Bonnoratom,  Krasinski} and  is unlikely 
to be extended to non-spherical systems \cite{non-extension, 
NolanPRD}. Many 
papers in the following years 
\cite{variousworks, NoerdlingerPetrosian, SatoMaeda, Sussman, 
Death, Ferrarisetal, CFV, Bombelli, Bonnoratom, Nolan, 
Nolan2, NolanPRD, Bushaetal, GaoZhang, SultanaDyer, Pioneer, 
NesserisPerivolaropoulos, Price,  BalagueraNowakowski, 
Mashhoonetal} presented 
contradictory results, casting a 
shadow of ambiguity on the problem (see \cite{reviews, 
Krasinski} for brief 
reviews). The most popular model consists of a test particle in 
a quasi-circular 
orbit around a Newtonian central object. It appears that many 
of the contradictory results are simply due to the use of 
different or unphysical coordinates \cite{CFV}. Moreover, a 
quantitative 
answer to the problem of how much the cosmic expansion affects 
local dynamics differs according to the type of local system 
considered. If the FLRW metric is an adequate model of spacetime 
down to small scales, then weakly gravitating systems of size 
small in comparison to the Hubble radius $H_0^{-1}$ do 
participate in the expansion, but the effect is so small to be 
completely negligible for practical purposes. When 
the size of the weakly gravitating system becomes  a larger 
fraction of the Hubble radius, the cosmic expansion plays a 
significant role in the dynamics; this is the situation of large 
scale structures \cite{NoerdlingerPetrosian, 
SatoMaeda, Bushaetal, BalagueraNowakowski}.

A recent paper by Price \cite{Price} studies a classical atom in 
a de Sitter background, with arbitrary strength of the coupling 
between an electron and the central charge. A new result is 
the ``all or nothing'' behaviour: if the coupling is weak the 
``atom'' is comoving with the rest of the universe, while if the 
coupling is very strong the ``atom'' is only slightly perturbed 
by a transient and  does not expand \cite{Price}. 
This work breaks free of the standard assumption of previous 
literature that the coupling (of a gravitationally, instead of 
electrically, bound system) is weak. However, it has 
two fundamental limitations: first, the cosmological 
background is restricted to be de Sitter space, which is very 
special: in fact, the de Sitter metric can be put in static 
form, which may explain why strongly bound systems in this 
background do not expand. Second, the classical atom is not an 
adequate model when the local energy density and stresses of 
the 
central charge grow and induce local deviations from the 
cosmological metric. In these situations exact solutions of the 
Einstein  equations are needed to describe both the relativistic 
central object  with a  strong local field and the surrounding 
universe. Solutions of this kind could also be useful in 
studying the evolution of primordial black holes 
\cite{primordialBHs} regarded as 
probes of the early universe \cite{HaradaCarrMaeda, 
SaidaHaradaMaeda}. New and existing exact solutions of this kind 
are 
studied in Secs.~3 and~4.

There are also more modern and perhaps more compelling 
motivations to study the effect of the cosmological expansion on 
local physics. It is now well known from the observations of 
supernovae of type Ia that the expansion of the universe is 
accelerated \cite{SN}. Marginal evidence for an equation of 
state parameter $w\equiv \frac{P}{\rho}<-1$  (where $\rho$ and 
$P$ are the energy density and pressure of the cosmic fluid, 
respectively)  has led theorists to take seriously into 
account the possibility  of  a Big Rip singularity at a finite 
time in the future \cite{BigRip}. Various authors have studied 
how local systems  (clusters, galaxies, stars, {\em etc.}) are 
teared apart as the Big Rip is approached 
\cite{BigRip,NesserisPerivolaropoulos}. In this situation the 
catastrophic cosmological expansion is not merely a perturbation 
of the local dynamics, but dominates it.

The current inability to explain away the Pioneer anomaly has 
led some authors to attribute it to the effect of the 
cosmological expansion, although this possibility seems to be 
ruled out \cite{Pioneer}. Finally, independent motivation for 
studying the interplay between local and cosmological dynamics 
comes from another problem of principle in general relativity. 
If  
a universe dominated by phantom dark energy, which violates all 
 the energy conditions and causes $w\equiv P/\rho$ to be less 
than $-1$, is heading toward the Big Rip and all bound systems 
are gradually ripped apart \cite{BigRip}, it is legitimate to 
ask what is the 
fate of the most strongly bound local object, namely a black 
hole. Does the strong local field resist the expansion (as 
suggested by extrapolating Price's  work \cite{Price}), or does 
the horizon expand and disappear, exposing the central 
singularity before the Big Rip is reached?  This would entail 
violation of  cosmic censorship in its cosmological 
formulation \cite{generalizedCosmicCensorship} and would 
constitute a (further) argument against phantom energy. An 
answer comes from Ref.~\cite{BDE}, in which accretion of a 
phantom test fluid onto a Schwarzschild black hole is studied. 
The results are extrapolated to a gravitating fluid and the 
conclusion is reached that the black hole decreases its mass due 
to the fact that the gravitating energy density accreted 
$P+\rho$ is negative. Accretion proceeds  until the horizon 
disappears together with 
the central singularity before the Big Rip is reached 
\cite{BDE}. Although the extrapolation from a test to a 
gravitating fluid is quite plausible, it would be preferable to 
base the conclusion  on an exact solution of the Einstein 
equations displaying accretion of cosmic fluid. A step in this 
direction is taken with new solutions presented here.

In this paper we examine various exact solutions representing 
strong field objects in a cosmological background: they include 
the McVittie metric \cite{McVittie}, the Schwarzschild-de Sitter 
black hole, the Nolan interior solution \cite{NIS}, a solution 
found recently by Sultana and Dyer \cite{SultanaDyer}, and new 
exact solutions that are perfectly comoving.
 It turns out that the ``all or nothing'' behaviour discovered 
by Price \cite{Price} persists in the Schwarzschild-de Sitter 
black hole but it is  a peculiarity of the  de Sitter background 
adopted and more general FLRW backgrounds do not allow for it.
Participation of  a local object, even strongly bound, to the 
cosmological expansion seems to be the general rule, a 
conclusion supported  
by various exact solutions. 

The plan of this paper is as follows: Sec.~2 studies a Newtonian 
quasi-circular orbit and the effect of the cosmic expansion upon 
it, making clear the peculiarity of de Sitter space even for 
this kind of systems. Sec.~3 examines known and new exact 
solutions, while Sec.~4 contains a discussion and the 
conclusions. We adopt the notations of Ref.~\cite{Wald}.

\section{A Newtonian object embedded in a FLRW universe: 
quasi-circular orbits}

In the literature, the most common line of approach to the 
problem of the effect of the cosmic expansion on local systems 
is to consider a spherical Newtonian object of mass $M $ and a 
test particle in a circular orbit of radius $r$ around it, and 
then ``switch on'' the cosmological dynamics as a small 
perturbation of this  two-body problem (elliptical orbits were 
considered in Ref.~\cite{Bombelli}). Motivated by 
the current model of our universe (and by simplicity), and 
following most authors, we consider as the background a 
spatially 
flat FLRW  universe 
described by the line element
\begin{equation}\label{1}
ds^2=-dt^2+a^2(t)\left( dx^2+dy^2 + dz^2 \right)
\end{equation}
in comoving coordinates. 
The evolution equation for the physical radial coordinate of the 
otherwise circular orbit is
\begin{equation} \label{2}
\ddot{r}=\frac{ \ddot{a}}{a}\, r -\frac{GM}{r^2} 
+\frac{L^2}{r^3} 
\;,
\end{equation}
where $L$ is the (constant) angular momentum per unit mass of 
the test particle and an overdot denotes differentiation with 
respect to comoving time. This equation is derived in several 
ways in 
the many papers on this subject: they range from heuristic 
derivations (e.g., \cite{variousworks,Price}) to calculations 
using the 
geodesic deviation equation in a locally inertial frame of the 
cosmological metric  (\ref{1}) \cite{footnote1, CFV, 
Mashhoonetal}. 
This 
derivation uses 
Fermi normal coordinates, regarded as the physical coordinates 
connected 
to a freely falling observer in the cosmological gravitational 
field. Other derivations of eq.~(\ref{2}) 
\cite{NoerdlingerPetrosian, NesserisPerivolaropoulos} use 
various 
approximations to equations for timelike geodesics in the 
McVittie metric \cite{McVittie}, in the limit in which a  
central object produces only a small deviation from the 
cosmological 
background. If the current era in the history of the universe is 
considered, the term $ \ddot{a} r/ a $ on the right hand 
side of eq.~(\ref{2}) is a small perturbation when 
$r<<H_0^{-1}$, where $H_0$ is the present value 
of the Hubble parameter $H \equiv \dot{a}/a$. This correction 
becomes increasingly 
important as the size of the orbit increases in comparison with 
the Hubble radius $ H_0^{-1}$, for example going 
to galaxy clusters and superclusters, for which it is 
certainly not 
negligible \cite{NoerdlingerPetrosian, 
SatoMaeda, Bushaetal, BalagueraNowakowski}.

Note that, for a decelerating universe, the term 
$\frac{\ddot{a}}{a}\, r$ is negative and is considered as such 
in pre-1998 literature, while it gives a 
positive contribution to $\ddot{r}$ in an accelerated universe, 
which is the case of the present epoch, as is now well known 
from 
the study of supernovae of type Ia at high redshift \cite{SN}.

Whether $ \ddot{a} r/ a $ can be treated as  
a perturbation or not, both the central object and the FLRW 
background are  assumed to be spherically symmetric, which 
yields conservation of the angular 
momentum per unit mass $L$ of the test particle,
\begin{equation}  \label{3}
r^2\dot{\varphi}=L 
\end{equation}
in spherical coordinates. 
This equation holds true in later sections in which we consider 
exact solutions of the Einstein equations describing a strongly 
gravitating, spherically symmetric, central object embedded in  
a FLRW universe. It follows that, if $r(t)$ increases with time, 
the angular velocity $\varphi(t)$ decreases. The test particle 
has to cover  a larger linear distance to attempt to close its 
orbit, which would be circular in the absence of the 
cosmological perturbation.

A different point of view is the one, found in recent  
literature, in which it is assumed that the universe is 
dominated by phantom energy with equation of state 
parameter $ w\equiv P/\rho<-1$ 
\cite{phantom}, and is heading toward a Big Rip in which the 
scale factor $a(t)$ diverges at a finite time $t_{rip}$.

A long time before the Big 
Rip, the term $\frac{\ddot{a}}{a}\, r$ can be treated as a small 
perturbation. In flat  space, at the Newtonian level, 
this term is absent, Kepler's third law yields 
$\dot{\varphi}^2 r^3=GM$, and the two terms $-GM/r^2$ and $ 
L^2/r^3$ in eq.~(\ref{2}) cancel each other, leading to circular 
orbits $r=$constant. When the perturbation 
$\frac{\ddot{a}}{a}\, r$ is introduced, this is no longer true 
and this perturbation changes the (otherwise circular) orbit. 
When the Big Rip is approached as $t\rightarrow t_{rip}^{-}$ and 
the central object is weakly gravitating (Newtonian), the term 
$\frac{\ddot{a}}{a}\, r$ dominates over the terms in $r^{-2}$ 
and $r^{-3}$, leaving 
\begin{equation}  \label{4}
\ddot{r}=\frac{\ddot{a}}{a}\, r 
\end{equation}
as the asymptotic evolution equation for the physical radius of 
the orbit. The solution of eq.~(\ref{4}) is $ r\propto a(t)$ or, 
the orbit becomes 
comoving with the cosmic substratum. For simplicity, assume that 
the phantom energy dominating the cosmic dynamics and causing 
the Big Rip has constant equation of state $P=w\rho$, where 
$w<-1$. Then, the scale factor is
\begin{equation} 
a(t)= a_0 \left(t_{rip} -t \right)^{\alpha} \;, \;\;\;\;\;\;\;
\alpha=\frac{2}{3\left( w+1 \right)}<0 \;
\end{equation}
and eq.~(\ref{4}) reduces to 
\begin{equation} \label{6}
\ddot{r}=\frac{ \alpha\left( \alpha-1\right)}{\left( t_{rip}-t 
\right)^2} \,\, r \;,
\end{equation}
which has the general solution
\begin{equation} \label{7}
r(t)=A  \left( t_{rip}-t\right)^{\alpha}+B \left( t_{rip}-t 
\right)^{1-\alpha} \;,
\end{equation}
where $A$ and $B$ are integration constants. Since 
$1-\alpha=\frac{3w+1}{3\left( w+1\right)}>0$, the second term on 
the right hand side of eq.~(\ref{7}) becomes negligible with  
respect to the first one as the Big Rip is approached. 
Therefore,  the solution (\ref{7}) reduces to  $r(t)\propto 
a(t)$ as $t\rightarrow 
t_{rip}^{-}$, i.e., the putative circular orbit becomes 
comoving.

The angular motion is obtained by integrating 
eq.~(\ref{3}), which yields
\begin{eqnarray}
\varphi(t) &= &\int dt 
\, \frac{L}{r^2}=\frac{3L(w+1)}{1-3w}\left[ 
A^2\left( t_{rip}-t\right)^{\frac{1-3w}{3\left( w+1\right)}} 
+AB \right]^{-1} \nonumber \\
&&\nonumber \\
&+& \varphi_0 \simeq 
\frac{3(w+1)}{1-3w}   \frac{L}{A^2}   \left( 
t_{rip}-t\right)^{\frac{3w-1}{3\left( w+1\right)}}+\varphi_0 \;.
\label{8}
\end{eqnarray}
Then $\varphi(t)\rightarrow \varphi_0$ as 
$t\rightarrow t_{rip}^{-}$: as the Big Rip is approached and $r$ 
grows without 
bound (but comoving), the angular motion slows down and freezes. 
Strictly speaking, the orbit is never ``disrupted'' before the 
spacelike singularity  is reached; it just 
participates in the cosmic expansion that is accelerating 
catastrophically. In this sense, it is not true that  bound 
systems become unbound: this is a rather misleading sentence 
often echoed by the media. If one wants to insist on the use of 
this terminology, the meaning of ``bound'' and ``unbound'' 
system 
should be clearly defined. For example, one may think of 
(arbitrarily) setting the threshold between ``bound'' and 
``unbound'' when, in eq.~(\ref{2}), the cosmological term 
$ \ddot{a} r/ a $ becomes of the order of the other two 
terms $-GM/r^2$ and $L^2/r^3$. The condition 
$ \ddot{a} r/ a \approx GM/r^2$ can be expressed by 
saying that the time scale $\tau \sim \sqrt{  
\frac{a}{\ddot{a}} }$ is of the order of the free fall time 
scale of  a fictitious region of radius $r$ containing the mass 
$M$, 
i.e., 
$ \frac{r}{v_E}\sim \sqrt{ \frac{r^3}{GM}}$, where $v_E$ is the 
escape velocity. Or, in other words, the energy density $\sim 
M/r^3$ of this fictitious region equals the cosmological density 
$\sim \ddot{a}/a $ which grows in a phantom-dominated universe. 
A 
more precise characterization of when a ``bound'' system is 
``disrupted'' is given  by 
Nesseris and Perivolaropoulos \cite{NesserisPerivolaropoulos}. 
These authors consider the effective potential $V(t,r)$ of the 
one-dimensional equation of motion of the test particle and 
determine when its minimum disappears by solving numerically 
the equation $\partial V/\partial r=0$ (the location of this 
minimum depends on time). This procedure corrects our order of 
magnitude estimate by a factor $\sim 3$. The effective potential 
used is the subject of the next subsection.

\subsection{Effective potential, Lagrangian, and Hamiltonian}

By rewriting eq.~(\ref{2}) as
\begin{equation} \label{9}
\ddot{r}=-\frac{GM}{r^2}+\frac{L^2}{r^3}+\frac{\ddot{a}}{a}\, r 
=-\frac{\partial V}{\partial r} \;,
\end{equation}
integrating with respect to $r$, and setting to zero an 
arbitrary integration function of time, one obtains the 
effective potential
\begin{equation} \label{10}
V\left( t, r \right)=- \frac{\ddot{a}}{2a}\, r^2
-\frac{GM}{r} +\frac{L^2}{2r^2} \;.
\end{equation}
In a general FLRW spacetime this effective potential for the  
bound system test particle-central object depends on time 
because of the cosmological term
$ -\frac{\ddot{a}}{2a}\, r^2$ and one can say that this system 
exchanges energy with the cosmological 
background. However, in the special case of  a de Sitter 
background described by the scale factor $a(t)=a_0 \exp( H_0 
t)$ with constant $H_0$, this term is time-independent and 
reduces to $ 
-H_0^2r^2/2$ --- no energy is then exchanged between the 
two-body 
system and the cosmological background, and the energy of the 
former is constant. This simplification was noted by Price 
\cite{Price}, who restricted his study of the effect of 
cosmological dynamics on local systems to a de Sitter 
background. 

Given the effective potential (\ref{10}), it is straightforward 
to deduce the effective Lagrangian
\begin{equation} \label{11}
{\cal L} \left( t, r, \dot{r} 
\right)=\frac{\dot{r}^2}{2}+\frac{GM}{r} 
-\frac{L^2}{2r^2}+\frac{\ddot{a}}{2a} 
\, r^2 \;,
\end{equation}
which reproduces eq.~(\ref{2}) through the Euler-Lagrange 
equation
\begin{equation} \label{12}
\frac{d}{dt}\left( \frac{\partial {\cal L}}{\partial \dot{r}} 
\right)-\frac{\partial {\cal L}}{\partial r}=0 \;,
\end{equation}
and the effective Hamiltonian
\begin{equation} \label{13}
{\cal H}\left( t, r, p_r \right)=
\frac{p_r^2}{2} -\frac{GM}{r} 
+\frac{L^2}{2r^2}-\frac{\ddot{a}}{2a} \, r^2 \;,
\end{equation}
where $p_r=\partial {\cal L}/\partial \dot{r}=\dot{r}$ and
\begin{equation} \label{14}
\dot{r}=\frac{\partial {\cal H}}{\partial p_r} \;, \;\;\;\;\;\;\
\dot{p}_r=-\frac{\partial {\cal H}}{\partial r} \;.
\end{equation} 
This Hamiltonian is, of course, time-dependent for any FLRW 
background that is not a de Sitter space:
\begin{equation} 
\frac{\partial {\cal H}}{\partial t}=
- \frac{\partial {\cal L}}{\partial t}=
\frac{\partial V}{\partial t}  \;.
\end{equation}
For a de Sitter background $a=a_0\mbox{e}^{H_0 t}$ with 
$H_0=\sqrt{\Lambda/3}$ (where $\Lambda >0 $ is the cosmological 
constant), the effective potential (\ref{10}) is 
consistent with the weak-field limit of the Schwarzschild-de 
Sitter (or K\"{o}ttler) metric, given by the line element
\begin{widetext}
\begin{equation} \label{15}
ds^2=-\left( 1-\frac{2GM}{r} 
-\frac{\Lambda r^2}{3} \right)dt^2
+ \left( 1-\frac{2GM}{r}-\frac{\Lambda r^2}{3}\right)^{-1}dr^2 
+r^2 d\Omega^2 \;,
\end{equation}
\end{widetext}
in static coordinates, where $d\Omega^2=d\theta^2 +\sin^2 
\theta \, d\varphi^2 $ is the line element on 
the unit 2-sphere.  The analogue of the Newtonian potential 
$\Phi_N$ for a test particle can be read off the $\left( 
0,0\right)$ component of the K\"{o}ttler metric
$g_{00}=-\left[ 1+2\Phi_N(r) \right]$, which yields
\begin{equation} \label{16}
\Phi_N(r)=-\frac{GM}{r}-\frac{\Lambda r^2}{6} \;.
\end{equation}
On the other hand, the cosmological part of the potential 
(\ref{10}) in the equation of motion (\ref{2}) of the test 
particle in the de Sitter background with $H_0=\sqrt{\Lambda/3}$ 
is
\begin{equation} 
-\frac{ \ddot{a}}{2a}\, r^2=-\frac{H_0^2 
r^2}{2}=-\frac{\Lambda r^2}{6} \;,
\end{equation}
which is consistent with eq.~(\ref{16}). This is not a 
coincidence because eq.~(\ref{2}) can be derived as a special 
approximation of the timelike geodesic equation of the McVittie 
metric \cite{McVittie}, which reduces to the K\"{o}ttler metric 
for a de Sitter background (see Sec.~3.3).

By adding the 
centrifugal potential term $\frac{L^2}{2r^2}$ one recovers the 
effective potential (\ref{10}) for the equivalent 
one-dimensional problem and $V(r)=\Phi_N(r)$.

\subsection{Orbits of constant radial coordinate}

It is time to comment on the physical meaning of the radial 
coordinate $r$. As shown in Ref.~\cite{CFV}, this coincides with 
the proper radius when $rH_0<<1$ and receives corrections of 
higher order when $r$ is larger and larger with respect to the 
Hubble radius $H_0^{-1}$. When radii  $r\sim H_0^{-1}$ are 
considered, $r$ assumes the meaning of comoving radial 
coordinate in the FLRW metric (\ref{1}). However, when the field 
of the central object is strong, neither of the above describes 
precisely the meaning of $r$ for  $r<< H_0^{-1}$.

The question of whether orbits of constant radial coordinate $r$ 
exist is of some interest. Orbits of constant radius are found 
by Bonnor \cite{Bonnoratom} who, similarly to Price 
\cite{Price},  considers a classical atom embedded in a 
cosmological background. Two situations can be distinguished.\\
i) The cosmological background is {\em not} a de Sitter space. 
Then, $\ddot{a}/a$ depends on time and, imposing $ 
r=r_0\equiv$constant, one obtains
\begin{equation} \label{18}
\frac{\ddot{a}}{a}\, 
r_0-\frac{GM}{r_0^2}+\frac{L^2}{r_0^3}=0 \;.
\end{equation}
The first term on the left hand side depends explicitly on 
time, 
while the remaining terms are time-independent, hence this 
equation can not be satisfied and orbits of constant $r$ do not 
exist in this case.\\
ii) The cosmological background is de Sitter space. Then 
imposing $r=r_0\equiv$constant yields the quartic  
equation for $r_0$
\begin{equation} \label{19}
H_0^2r_0^4 -GMr_0+L^2=0 \;.
\end{equation}
This equation can, in principle, have real solutions under  
conditions which are discussed in the next subsection.\\

\subsection{Test particle around a Newtonian central object in a 
de Sitter background: phase plane analysis}

The equation of motion of an electron in a classical atom  
embedded in a de Sitter background is analogous to our 
eq.~(\ref{2})  and is solved numerically by Price \cite{Price}. 
Contrary to previous authors, Price does 
not restrict himself to considering weak couplings of the 
electron, 
which is the equivalent of the situation considered so far in 
our paper, of a test particle in the field of a Newtonian, 
weakly gravitating central object embedded in a cosmological 
background. Price considers instead arbitrarily strong coupling 
of the electron to the central charge and discovers an ``all or 
nothing'' behaviour: if the coupling is weak the electron 
trajectory becomes comoving with the de Sitter substratum while, 
if the coupling is strong, the evolution of the orbit exhibits a 
transient after which it is essentially unperturbed. A critical 
value of the angular momentum separates these two behaviours. 
This ``all or nothing'' feature went undetected in the 
(abundant) previous literature  which did not break free of the 
weak coupling assumption. In our formalism, allowing for an 
arbitrarily strong gravitational field due to the central object 
means making this object relativistic and one must leave the 
regime in which the latter is merely a perturbation of a cosmic 
substratum, and move to a fully relativistic regime by studying 
exact solutions describing a strongly gravitating central 
object embedded in a FLRW background. This will be done in 
the next section. Here we limit ourselves to weak 
coupling and we provide a phase space analysis of the equation 
of motion analyzed numerically by Price \cite{Price}, in the 
analogous situation of a two-body system completely ruled by 
gravity. This phase space analysis is still missing in the 
literature.

By introducing variables $x\equiv r$ and $y\equiv \dot{r}$ (not 
to be confused with Cartesian coordinates in the de Sitter 
background (\ref{1})), eq.~(\ref{2}) is written as the 
autonomous dynamical system
\begin{eqnarray}
\dot{x}&=& y \;, \label{20}\\
&&\nonumber \\
\dot{y}&=& H_0^2 x -\frac{GM}{x^2}+\frac{L^2}{x^3} \;,\label{21}
\end{eqnarray}
for $x>0$ and any real value of $y$. We assume $L\neq 0$ (the 
case $L=0$ will be discussed later). Equilibrium points, if they 
exist, correspond to orbits of constant radius $\left( x,y 
\right)=\left( r_0, 0 \right)$ of the kind discovered by Bonnor 
\cite{Bonnoratom}. 
The search for these fixed points is equivalent to solving 
eq.~(\ref{19}), or
\begin{equation} \label{22}
\psi (x)\equiv H_0^2x^4-GMx+L^2=0 \;.
\end{equation}
The function $\psi(x)$ is represented by a quartic parabola with 
its concavity facing upward, therefore there can be solutions 
of eq.~(\ref{22}) only if the minimum $\psi_{min}$ of $\psi(x)$ 
is non-positive. If $\psi_{min}=0$ there are two coincident 
roots, while if $\psi_{min}<0$ there are two real distinct 
roots, and no real roots exist if $\psi_{min}>0$. The 
study of the first derivative $d\psi/dx$ 
establishes that $\psi(x)$ has the absolute minimum
\begin{equation} \label{23}
\psi_{min} \equiv \psi\left(x_{min} \right)=L^2- 
\frac{3z}{H_0^2}
\end{equation}
at
\begin{equation}  \label{24}
x_{min}=\left( \frac{GM}{4H_0^2} \right)^{1/3} \;,
\end{equation}
where 
\begin{equation} \label{25}
z\equiv  \left( \frac{GMH_0}{4 } \right)^{4/3} 
\end{equation}
is a dimensionless variable. It follows that 
$\psi_{min}>0$, hence there are no orbits of constant $r$ if 
\begin{equation} 
 L> \sqrt{3} \, \left( \frac{GM}{4\sqrt{H_0}} 
\right)^{2/3}\equiv L_c \;.
\end{equation}
There is a single orbit of constant $r=x_{min}$ if $L=L_c$, and 
there are two orbits of constant radii $r_{1,2}$ with 
$r_1<x_{min}<r_2$ when $L<L_c$. These are all the fixed points 
$\left( x_0, 0 \right)$ of the dynamical system  (\ref{20}) and 
(\ref{21}). In order to assess the stability of these fixed 
points let us consider perturbations described by 
$x(t)=x_0+\delta x(t)$, $y(t)=\delta y(t)$. Eqs.~(\ref{20}) and 
(\ref{21}) yield the evolution equation for the orbital radius
\begin{equation} \label{26}
\delta \ddot{r}+\omega^2 \delta r\simeq 0 \, ,\;\;\;\;\;\;\;\;
\omega^2=\frac{L^2}{r^4}-3H_0^2 \;.
\end{equation}
Linear stability corresponds to $\omega^2\geq 0$. Orbits of 
constant 
$r$ 
exist when $L\leq L_c$, with the equality corresponding to a 
single orbit of radius $r_0$ and the strict inequality to two 
orbits of radii $r_{1,2}$ with  $r_1<x_{min}<r_2$. The  
stability condition of an 
orbit of constant  radius $r_{*}$ is equivalent to $ r_*\leq 
\sqrt{  \frac{L}{\sqrt{3} \, H_0}} $. In 
the 
case of a single orbit it is $L=L_c$ and $r_*=r_0=x_{min}\leq 
x_{min} $ and the stability condition is satisfied. In the case 
of two 
distinct orbits it is $r_1<x_{min} <r_2$ and therefore the outer 
orbit is unstable. The inner orbit is stable only if $r_1\leq 
\sqrt{ \frac{L}{\sqrt{3} \, H_0}}$. Since $\psi\left( 
x=\sqrt{\frac{L}{\sqrt{3}\, H_0}}\right)=\frac{4L^2}{3}-GM 
\sqrt{ \frac{L}{\sqrt{3}\, H_0}} <0$ when $L>L_c$ and $\psi(x)$ 
is a decreasing function between $x=0$ and $x_{min}$, while 
$\psi(r_1)=0$, it must be $r_1< \sqrt{\frac{L}{\sqrt{3} \, 
H_0}}$ and therefore the inner orbit is stable.

At large values of $x$, eq.~(\ref{21}) reduces to the asymptotic 
equation $\dot{y} \simeq H_0^2 x$, which has the solution 
\begin{equation} 
\left( x(t),  y(t) \right)=\left( x_* \mbox{e}^{H_0 t}, H_0x_* 
\mbox{e}^{H_0 t } \right)
\end{equation} 
represented by the line $y=H_0 x$ in the $\left( x, y 
\right)$ plane. This solution is an attractor for large values 
of $x$. In fact, by perturbing the solution as described by $ 
x(t)=x_* \mbox{e}^{H_0 t} +\delta x(t)$ and $y(t)=H_0 x_* 
\mbox{e}^{H_0 t} +\delta y(t)$ one obtains, to first order in 
the perturbations, $\delta \ddot{x}=H_0^2 \delta x + \, ...$ 
with solution $\delta x(t)=\delta_0 \mbox{e}^{H_0 t}$ (with 
$\delta_0$ a constant). Therefore, the ratios
\begin{equation} 
\frac{ \delta x(t)}{x_* \mbox{e}^{H_0 t} } 
=\frac{\delta_0}{x_0} \;, \;\;\;\;\;\;\;\;\;\;
\frac{ \delta y(t)}{H_0 x_* \mbox{e}^{H_0 t} } 
=\frac{\delta_0}{x_0} 
\end{equation}
stay small if they start small ($\left| \frac{\delta_0}{x_0} 
\right|<<1$).

The qualitative picture of the phase space can be completed as 
follows. The system (\ref{20}) and (\ref{21}) has the first 
integral
\begin{equation} \label{energy}
E=\frac{y^2}{2}-\frac{H_0^2 
x^2}{2}+\frac{L^2}{2x^2}-\frac{GM}{x}
\end{equation}
corresponding to the energy of the test particle, which is 
conserved in a de Sitter background.\\\\
\noindent $\bullet$~~ If $L>L_c$ there are no fixed points and 
$y=H_0 x$ is an attractor for large values of $x$. Then 
$\dot{y}=\psi(x)/x^3 >0 $ (the equation $\psi=0$ has no real 
roots for 
$L>L_c$). Hence $y=\dot{x}$ always increases and either 
$y\rightarrow +\infty$ or $y(t)$ has a horizontal asymptote with 
$y(t)\rightarrow y_0$ as $t\rightarrow +\infty$. If 
$y=\dot{x}\rightarrow +\infty$, then either $x(t) \rightarrow 
+\infty$ or $x(t)$ has  a vertical asymptote with $x\rightarrow 
x_0$ and $\dot{x}\rightarrow +\infty$. There are, in principle, 
four possibilities.\\
1) $x(t) \rightarrow +\infty$ and $ y(t)\rightarrow +\infty$; 
then the orbit of the solution is necessarily captured in the 
attraction basin of the attractor $y=H_0x$.\\
2) $x(t) \rightarrow +\infty$ and $ y(t)\rightarrow y_0$; this 
is not possible because as $x\rightarrow +\infty$ it is $y=H_0 x 
\rightarrow +\infty$.\\
3) $x(t) \rightarrow x_0 $ and $ y(t)\rightarrow +\infty$ 
($x(t)$ has  a vertical asymptote $\dot{x}\rightarrow +\infty$); 
then 
the energy $E$ given by eq.~(\ref{energy}) diverges at late 
times,  
which is not possible because $E$ is constant.\\
4) $x(t) \rightarrow x_0 $ and $ y(t)\rightarrow y_0 $; this 
means that there is an attractor point, then $\dot{y}=H_0^2 x 
-GMx^{-2}+L^2 x^{-3} \rightarrow 0 $ as $t\rightarrow +\infty$ 
then $x_0$ must be a root of the equation $\dot{y}=0$ but this 
is not possible because this equation  has no real roots when 
$L>L_c$. This case is not possible.  In summary, for $L = L_c$ 
all 
the trajectories of the solutions in phase space go to the 
attractor $y=H_0x$ as  $t\rightarrow +\infty$, i.e., all 
physical orbits become comoving.\\\\
\noindent $\bullet$~~ If $L=L_c$ there are the attractor 
$y=H_0x$ as $x\rightarrow +\infty$ and the fixed point 
$x_0=x_{min}=\left( \frac{GM}{4H_0^2} \right)^{1/3}, y_0=0$ 
which is the only real root of the equation $\dot{y}=0$. Since 
there is only one such root, {\em a priori} either $\dot{y}>0$ 
or $\dot{y}<0$ 
along any orbit that does not coincide with the fixed point, 
without the possibility of changing sign during the evolution. 
However, we know that $\dot{y}=\psi(x)/x^3>0 \; \,\forall x\neq 
x_0$ 
then $y=\dot{x}$ is always increasing for $x\neq x_0$. By 
repeating the reasoning of the case $L>L_c$, we have now two 
possibilities:\\
1) $x(t) \rightarrow +\infty $ and $ y(t)\rightarrow 
+\infty$, in which case the orbit of the solution in phase 
space gets captured by the attractor 
$y=H_0x$ at infinity.\\
2) $ \left( x(t), y(t) \right)  \rightarrow \left( x_0 , 0 
\right) $  with $y(t)\rightarrow 0^{-}$ so $y=\dot{x}<0$ and 
$x(t)$ is always decreasing to $x_0$. Then, $\ddot{x}>0$ with 
$\dot{x}<0$; the point $\left( x_0 , 0 \right)$ is an attractor 
point.  Since $\dot{y}>0 \, \forall \left( x,y \right)\neq 
\left( x_0, y_0 
\right) $ there are no periodic orbits. The cases in which 
$\left( x(t), y(t) \right) \rightarrow \left( +\infty, y_0 
\right) $ or   
$\left( x(t), y(t) \right) \rightarrow \left( x_0, +\infty 
\right) $   are excluded as previously  discussed for $L>L_c$. 
In summary, for $L=L_c$ the orbits of the solutions either go to 
the attractor point or to the attractor at infinity 
$y=H_0x$.\\\\
\noindent $\bullet$~~ If $0< L<L_c$  there are the 
equilibrium points $\left( r_{1,2}, 0 \right)$ and the attractor 
at infinity $y=H_0x$ with $r_1<x_{min}<r_2$. There are two real 
distinct  roots of the equation $\dot{y}=\psi(x)/x^3=0$ and it 
is   
$\dot{y}>0$ for $x<r_1$ and for $x > r_2$, while it is 
$\dot{y}<0$ for $r_1<x<r_2$.

For $x<r_1$ the function $y(t)=\dot{x}$ 
is increasing and either $y\rightarrow +\infty$ or 
$y(t)\rightarrow y_0^{-}$ (horizontal asymptote of $y(t)$). If 
$y\rightarrow +\infty$ it must be because $x\rightarrow 0$, the 
familiar situation in which the Newtonian centrifugal potential 
$ L^2 x^{-3}$  dominates and repels the particle to larger $x$. 
If $y(t)$ has a horizontal asymptote, $y\rightarrow y_0$, then 
$\dot{y}\rightarrow 0$ and the attractor point $\left( r_1,0 
\right)$ is approached.\\
For $ r_1 <x<r_2$ it is $\dot{y}=\psi(x)/x^3 <0$ so $y(t)$ is 
always decreasing and either $y(t)\rightarrow -\infty$ or $y(t)$ 
has  a horizontal asymptote $y\rightarrow y_0^{-}$. The first 
situation is not possible because it would imply that the 
energy $E$ diverges, while it is instead forced to be constant. 
Therefore, $y(t)$ must have a horizontal asymptote and 
$\dot{y}\rightarrow 0$. Then the attractor point (zero of the 
function $\psi(x)$) is approached.\\
For $x>r_2$ it is $\dot{y}>0$ and either $y(t)\rightarrow 
+\infty$ or $y(t)\rightarrow y_0^{+}$ (horizontal asymptote). If 
$y\rightarrow +\infty$ there are, in principle, two 
possibilities:\\
1) $ x(t)\rightarrow +\infty$ and $y(t)\rightarrow +\infty$; 
then the orbits  of the solutions are captured by the attraction 
basin of $y=H_0x$.\\
2) $x(t)$ stays finite and $y(t)\rightarrow +\infty$. Again, 
this would give $E\rightarrow +\infty$, which is excluded.

\subsubsection{Zero angular momentum} 

Finally, let us consider the special case $L=0$ in which the 
radial coordinate of the test particle satisfies
\begin{equation}  \label{29}
\ddot{r}=H_0^2 r -\frac{GM}{r^2} \;.
\end{equation}
Only one ``orbit'' of 
constant physical radius $r_0=\left( 
GM/H_0^2\right)^{1/3}$ exists in this case. This situation is 
not possible 
in flat space and 
it corresponds to the cosmological ``force'' $H_0^2r$ exactly 
compensating the 
attractive  force of the central object $-GM/r^2$. As is clear 
from the 
physical point of view, this position of equilibrium is 
unstable because an arbitrarily small radial displacement will 
move the test particle to a region where one of the two forces 
is dominant. The value $r_0 =\left( GM/H_0^2 \right)^{1/3}$ 
corresponds to a maximum of the effective potential 
\begin{equation} 
V(r)=-\frac{H_0^2r^2}{2}-\frac{GM}{r} \;.
\end{equation}
In fact, setting $dV/dr=0$  reproduces this value of $r$ and 
$ d^2V/dr^2\left.\right|_{r_0}=-3H_0^2<0$. This simple solution 
is apparently missed in previous literature.

\subsection{Purely radial motion in  a general FLRW background}

One can generalize to any FLRW background the search for 
solutions with purely 
radial motion satisfying
\begin{equation} \label{30}
\ddot{r}=\frac{\ddot{a}}{a}\, r-\frac{GM}{r^2} \;.
\end{equation}
Orbits of constant $r$ are impossible if the background is not 
de Sitter space. We consider universes with  scale 
factor given by a power-law $a(t)=a_0 \, t^p$ (with $p$ a 
constant). 
Then, it is easy to find the power-law solution 
\begin{equation} \label{31}
r_*(t)=\left[ 
\frac{GM}{p\left(p-1\right)+\frac{2}{9}}\right]^{1/3} 
t^{2/3}\equiv r_c t^{2/3} \;.
\end{equation}
This solution exists only for values of the constant $p$ 
satisfying $p<1/3$ or $p>2/3$, which follows from the 
requirement 
that the denominator of the constant $r_c $ in (\ref{31}) be 
positive. The  solution (\ref{31}) is unstable with respect to 
small radial 
perturbations, In fact, if $r(t)=r_*(t)+\delta r(t)$, using 
eq.~(\ref{31}) one finds the evolution equation for the linear  
perturbations $\delta r$
\begin{equation} \label{32}
\delta \ddot{r}-\left[ 3p\left( p-1\right)+\frac{4}{9} \right] 
\frac{\delta{r}}{t^2} =0 \;,
\end{equation}
which has solutions $\delta r(t)=\delta_0 \, t^{\beta}$ with 
$\delta_0$ a constant and 
\begin{equation} \label{33}
\beta_{\pm}=\frac{1}{2}\left( 1\pm \sqrt{ 12p\left( p-1 
\right)+\frac{25}{9}} \, \right) \;.
\end{equation}
In the allowed range of values of $p$, the discriminant $\Delta 
=12p\left( p-1 \right)+\frac{25}{9}$ satisfies 
$\sqrt{\Delta}>1/3$, which yields 
$\beta_{+}=\left( 1+\sqrt{\Delta} \right)/2 >2/3$ and the 
corresponding 
mode $\delta r_{+}(t) $ satisfies
\begin{equation} \label{34}
\frac{ \delta r_{+}(t) }{r_*(t)} \simeq 
t^{\beta_{+}-\frac{2}{3}} 
\;.
\end{equation}
Because of the positive exponent $\beta_{+}-2/3$, this ratio 
grows when $t\rightarrow +\infty$ and therefore the solution 
(\ref{31})  is unstable.

To conclude, one recognizes that the de Sitter universe, 
although simpler to study than a general FLRW background, is a 
very special case. Furthermore, 
the simple problem of a test particle in a  circular orbit is 
extremely simplified and perhaps is not the best physical system 
in which to study the competition between local attraction and 
cosmological expansion. The next step would be to consider the 
expansion of a  Newtonian star embedded in a FLRW universe. We 
skip this step and, in the next section, we consider instead a 
{\em 
relativistic} star embedded in a FLRW universe, and its 
Newtonian limit.

\section{A strongly gravitating object in a FLRW background}

The study of a classical atom in a de Sitter background by 
Price \cite{Price} shows  that systems that are strongly bound 
``resist'' the cosmological  expansion and are only perturbed a 
little by it. This leads one to believe that this situation 
would carry over to the case of a gravitationally bound system 
and that a strongly gravitating central object, such as a black 
hole, would not be perturbed at all by the cosmic expansion. 
Whether this is true or not is  
the first question addressed in this section. Second, one would 
like to know whether this time-independent behaviour of a 
strongly gravitating central object (if it really carries over 
from Price's classical atom to the gravitating system) is 
peculiar of de Sitter space, 
or if it is valid in {\em general} FLRW backgrounds. After all, 
de 
Sitter space (other than the trivial Minkowski space with 
$a\equiv 1$) is the closest to a static space, and (a portion 
of) it can be 
expressed in static coordinates; therefore, it could be too 
special to derive general conclusions. Third, it is claimed 
in the literature that 
in a universe approaching the Big Rip all bound objects 
(galaxies, 
stars, atoms, {\em etc.}) are ripped apart before the 
singularity is reached \cite{BigRip}. What about a black 
hole horizon? If the expansion to a Big Rip tears apart a black 
hole horizon and a naked singularity appears, 
the cosmological version of cosmic censorship 
\cite{generalizedCosmicCensorship} is violated, 
the implications for black hole thermodynamics are 
non-trivial, and  the phantom energy causing the Big Rip 
would be questioned further from the point of view of 
fundamental physics. Regarding this problem, 
it is claimed that a black hole 
accreting phantom energy with $P<-\rho$ in a universe 
approaching the Big Rip {\em decreases} its mass and the horizon 
disappears together with the central singularity before the Big 
Rip is reached \cite{BDE}. This phenomenon avoids the violation 
of cosmic censorship by eliminating the central singularity 
altogether. New exact solutions can help clarifying this issue.

In all the situations mentioned above one can heuristically see 
the ``all or  nothing behaviour'' as resulting from the 
competition between 
two strong fields, the cosmological one and the local one due to 
the 
central object. It is conceivable that one of the two is locally 
stronger than the other and ``wins'' but, on the other hand, 
many examples of black holes are known which, placed in a 
strong external gravitational, electric, or magnetic field, have 
their horizons stretched  
\cite{KernsWild, FarooshZimmerman, BiniCherubiniMashhoon, 
DhurandharDadhich, Poisson, Anninosetal, Bretonetal, 
PatelTrivedi, Death, 
Doroshkevichetal, Ernst, ErnstWild, 
WildKerns, KulkarniDadhich, Krori, Bicak, FrolovNovikov}.

In this section we want to 
give a more precise meaning to these naive considerations, and  
a general picture of these phenomena that is not limited to a de 
Sitter background space. Exact 
solutions of 
the Einstein equations representing strongly gravitating objects 
embedded in a FLRW universe are needed. We neglect 
semiclassical Hawking radiation in the following.

\subsection{The McVittie solution}

An obvious starting 
point is the 
McVittie metric introduced in 1933 with the explicit purpose of 
investigating the effects of the cosmic expansion on local 
systems 
\cite{McVittie}. The line element is
\begin{equation} \label{McVittie}
ds^2=-\frac{  \left(1-\frac{m(t)}{2\bar{r}} \right)^2}{
\left(1+\frac{m(t)}{2\bar{r}} \right)^2} \, dt^2+
a^2(t) \left( 1+\frac{m(t)}{2\bar{r}} \right)^4 \left( 
d\bar{r}^2 +\bar{r}^2 d\Omega^2 \right)
\end{equation}
in isotropic coordinates, where the function $m(t)$ of 
the comoving time satisfies the equation \cite{McVittie}
\begin{equation}  \label{35}
\frac{\dot{m}}{m}=-\frac{\dot{a}}{a} 
\end{equation}
which yields
\begin{equation} \label{36}
m(t)=\frac{m_H}{a(t)} \;,
\end{equation}
where $m_H $ is a constant representing the Hawking-Hayward 
quasi-local mass \cite{Hawking,Hayward} (see 
Refs.~\cite{NolanPRD,Nolan} for a discussion). 
Eq.~(\ref{35}) was imposed 
by McVittie to explicitly forbid the accretion of cosmic fluid 
onto the central object (assumption {\em e)} of 
\cite{McVittie}). It 
corresponds to $ G_0^1=0$, which in turn implies that the 
stress-energy tensor component $T_0^1=0$ and there is no radial 
flow. In modern language, eq.~(\ref{35}) corresponds to the 
constancy of the Hawking-Hayward mass, $\dot{m}_H=0$.
It is important 
to recognize $m_H$ as the physically relevant mass 
(eventually related to 
the physical size of the horizon or of the central object) in 
order to avoid making  coordinate-dependent statements on 
the mass and size (cf., e.g.,  Refs.~\cite{GaoZhang, 
McClureDyerCQG}), or temperature \cite{SaidaHaradaMaeda} of the 
central object. 
$m(t)$ is just a metric coefficient 
in a particular coordinate system.

While there is little doubt that the McVittie metric represents 
some kind of strongly gravitating central object, its physical 
interpretation is not  completely clear  and is still debated 
today 
\cite{Sussman, Ferrarisetal, Nolan, Nolan2}. This 
metric reduces to the Schwarzschild solution in isotropic 
coordinates when $ a \equiv1$ and to the FLRW metric if $m 
\equiv 0$. 
However,  in general, the metric (\ref{McVittie})  can not be 
interpreted as describing a black hole embedded in a FLRW 
universe because it is 
singular on the 2-sphere $\bar{r}=m/2$ (which reduces to the 
Schwarzschild horizon if $a \equiv 1$)  
\cite{Ferrarisetal, Nolan,Sussman} and this singularity is 
spacelike \cite{Nolan,Nolan2}. It is claimed that the 
McVittie metric describes a point mass located at $\bar{r}=0$ 
(which is another, expected, singularity of the metric) embedded 
in a FLRW universe. However, this point mass is, in general, 
surrounded by the singularity at $\bar{r}=m/2$, which is 
difficult to interpret.  This 
singularity was studied in 
Refs.~\cite{Sussman,Ferrarisetal, Nolan}. Nolan 
\cite{NolanPRD} showed that 
this is a weak singularity in the sense that the volume of an 
object falling onto the $\bar{r}=m/2$ surface is not crushed to 
zero, and therefore the energy density of the surrounding fluid 
is finite. However, the pressure 
\begin{equation} \label{pressure}
P=-\, \frac{1}{8\pi G} \left[ 3H^2+\frac{2\dot{H}\left( 
1+\frac{m}{2\bar{r}} \right)  }{1-\frac{m}{2\bar{r}} }\right]
\end{equation}
diverges at $\bar{r}=m/2$ together with  the 
Ricci 
scalar $R=8\pi G \left( 3P-\rho \right)$ 
\cite{Sussman,Ferrarisetal, Nolan, McClureDyerCQG}. 
Notwithstanding this,  there is a situation in 
which the singularity disappears and the surface $\bar{r}=m/2$ 
describes a true black hole horizon: this happens when the 
background FLRW is de Sitter space.  Then $\dot{H}=0$ and the 
second term on the right hand side of the expression 
(\ref{pressure})---the only one causing $P$ to diverge---is 
absent   (this point was also noted in  Ref.~\cite{Nolan}).
A possible reason for the disappearance  of the singular surface 
in the 
de Sitter case is discussed below. Similar problems affect the 
charged McVittie metric \cite{GaoZhang} and the  
solutions of Thakurta \cite{Thakurta}, Vaidya \cite{Vaidya},   
Patel and Trivedi \cite{PatelTrivedi} representing rotating 
black hole-like objects in a cosmological background 
(\cite{McClureDyerCQG}---see also 
Refs.~\cite{Krasinski,McClureDyerGRG}).

\subsection{The Nolan interior solution: a relativistic star in 
a FLRW universe and its Newtonian limit}

It is of interest to study the behaviour of a relativistic star 
embedded in a FLRW background with respect to the problem of 
local physics versus cosmological expansion. The Nolan interior 
solution \cite{NIS} describes  a relativistic star of uniform 
density in such a background. The metric is
\begin{widetext}
\begin{equation}  \label{NIS}
ds^2=-\left[ \frac{ 1-\frac{m}{\bar{r}_0 }
+ \frac{m\bar{r}^2}{\bar{r}_0^3} \left(1-\frac{m}{4\bar{r}_0} 
\right) }{
\left( 1+\frac{m}{2\bar{r}_0} \right) \left( 1+\frac{ m 
\bar{r}^2}{2\bar{r}_0^3 } \right) } \right]^2 dt^2
+ a^2(t) \, \frac{ \left( 1+ \frac{m}{2\bar{r}_0}\right)^6 }{ 
\left( 1 +\frac{ m\bar{r}^2}{2\bar{r}_0^3} \right)^2 }
\left( d\bar{r}^2 +\bar{r}^2 d\Omega^2 \right) 
\end{equation}\end{widetext}
in isotropic coordinates, where $\bar{r}_0$ is the star radius, 
$\frac{\dot{m}}{m}=-\frac{\dot{a}}{a}$ (the condition forbidding 
accretion onto the star surface), and $0\leq \bar{r}\leq 
\bar{r}_0$. The interior metric is regular at the centre and is 
matched to the exterior 
McVittie metric at $\bar{r}=\bar{r}_0$ by imposing the 
Darmois-Israel junction conditions. The energy density is 
uniform and discontinuous at the surface $\bar{r}=\bar{r}_0$, 
while the pressure is continuous. These quantities  are given 
by \cite{NIS}
\begin{eqnarray}
\rho(t) &= & \frac{1}{8\pi G} \left[3H^2 +\frac{6m}{a^2 
\bar{r}_0^3 \left( 
1+\frac{m}{2\bar{r}_0 }\right)^6 } \right] \;, \label{birba1} \\
&& \nonumber \\
P\left( t, \bar{r} \right)&=& \frac{1}{8\pi G} \left[ -3H^2 
-2\dot{H}\, 
\frac{ \left(  1+\frac{m}{2\bar{r}_0 } \right) 
\left( 1 + \frac{m \bar{r}^2}{ 2\bar{r}^3_0 } \right) }{ 
1-\frac{m}{ \bar{r}_0} + 
\left( 1 - \frac{m}{ 4\bar{r}_0} \right) \frac{m 
\bar{r}^2}{ \bar{r}_0^3} } \right. \nonumber \\
&& \nonumber \\
&& \left. + \frac{1}{a^2} \frac{
\frac{3m^2}{\bar{r}_0^4} \left(  1-\frac{\bar{r}^2}{\bar{r}_0^2} 
\right)}{  \left( 1+ \frac{m}{2\bar{r}_0} \right)^6 \left[ 
1-\frac{m}{\bar{r}_0} + \left( 1-\frac{m}{4\bar{r}_0} \right) 
\frac{ m\bar{r}^2 }{\bar{r}_0^3 } \right]} \right] \;.\nonumber 
\\
&& 
\end{eqnarray}
The Nolan interior solution is a special member of 
Kustaanheimo's family 
of shear-free solutions \cite{Kustaanheimo} that generalizes the 
Schwarzschild 
interior solution of uniform constant density to the case of a 
time-dependent cosmological background. By setting $a\equiv 1$ 
one recovers familiar expressions for the Schwarzschild interior 
solution \cite{Wald}. The energy density is always positive and 
the condition $P\geq 0$ imposed by Nolan coincides with  
$\ddot{a}+3\dot{a}^2/2<0$ \cite{NIS}.

Let $ \Sigma_0(t)=\left\{ \left( t, \bar{r}, \theta, \varphi 
\right):\;\; \bar{r}=\bar{r}_0\right\} $ be the surface of the 
star at time $t$; by construction, the metric on this 2-sphere 
coincides with the McVittie metric
\begin{equation} 
ds^2\left.\right|_{\Sigma_0}=- \frac{ \left( 
1-\frac{m(t)}{2\bar{r}_0} \right)^2}{
\left( 1+\frac{m(t)}{2\bar{r}_0} \right)^2}\, dt^2 +a^2(t)\left(  
1+\frac{m(t)}{2\bar{r}_0} \right)^4 \bar{r}_0^2 d\Omega^2 \;.
\end{equation}
The proper area of $\Sigma_0$ is
\begin{equation} 
{\cal A}_{\Sigma_0}(t)=\int\int_{\Sigma_0} d\theta d\varphi \, 
\sqrt{ 
g_{\Sigma_0}}=4\pi a^2(t) \, \bar{r}_0^2 
\left( 1+\frac{m(t)}{2\bar{r}_0} \right)^4 \;,
\end{equation}
where $ g_{ab}\left.\right|_{( \Sigma_0) }$ is the metric on 
$\Sigma_0$ at a 
fixed time $t$ and $ g_{\Sigma_0}$ is its determinant. By using 
the Schwarzschild curvature coordinate $r\equiv \bar{r} \left( 
1+\frac{m}{ 2\bar{r}} \right)^2 $, one has
\begin{equation} 
{\cal A}_{\Sigma_0}(t)=4\pi a^2(t) \, r_0^2 \;.
\end{equation}
The star surface is comoving with the cosmic 
substratum and the proper curvature radius of the star 
is 
$r_{phys}(t)=a(t) \, \bar{r}_0 \left( 1+\frac{m}{2\bar{r}_0} 
\right)^2 $. Therefore, we have a local relativistic object with 
strong field which is perfectly comoving at all times: in this 
case the 
cosmic expansion ``wins'' over the local dynamics.

It is interesting to compute the generalized 
Tolman-Oppenheimer-Volkoff equation \cite{Wald} valid for 
this crude star model and to derive the first order correction 
to the  Newtonian equation of hydrostatic equilibrium.  For a 
perfect fluid described by the stress-energy 
tensor $T_{ab}=\left( P+\rho \right) u_a u_b +P g_{ab}$, the 
covariant conservation equation $\nabla^b T_{ab}=0$ splits into 
the two equations \cite{Wald}
\begin{eqnarray}
&& u^c \, \nabla_c \rho+ \left( P+\rho \right) \nabla^c u_c=0 
\;, 
\label{1001}\\
&& \nonumber \\
&& {h^c}_a \partial_c P +\left( P+\rho \right) u^c \, \nabla_c 
u_a=0 
\;,\label{1002}
\end{eqnarray}
where $u^a$ is the fluid four-velocity and $h_{ab}\equiv 
g_{ab}+u_a u_b$ defines the projector ${h_a}^c $ onto the 
3-space orthogonal to $u^a$. In comoving coordinates  it is 
$u^c  \propto \delta^{0c}$ and the normalization $u^c u_c=-1$ 
yields
\begin{equation} 
u^c=u \, \delta^{0c}=
\frac{ \left( 1+\frac{m}{2\bar{r}_0} \right) 
\left( 1+\frac{m \bar{r}^2 }{ 2\bar{r}_0^3} \right)}{
1 - \frac{m}{\bar{r}_0}  + 
\left( 1 - \frac{m}{4\bar{r}_0} \right) 
\frac{m\bar{r}^2}{\bar{r}_0^3} } \cdot \left( 1,0,0,0 \right) 
\;,
\end{equation}
or $u=\left| g_{00} \right|^{-1/2}$. Eqs.~(\ref{1001}) and 
(\ref{35}) then yield
\begin{widetext}
\begin{equation} 
\frac{\partial \rho}{\partial t}+3H \left( P+\rho 
\right)    
\left\{ 1- \frac{m}{\bar{r}_0} \left[
\frac{3}{2 \left( 1+\frac{m}{2\bar{r}_0} \right)} - 
\frac{ \bar{r}^2}{ 2\bar{r}_0^2}{ 
\left( 1+\frac{m\bar{r}^2 }{2\bar{r}_0^3} \right)^{-1} } \right] 
\right\} =0 \;, \label{1003}
\end{equation}\end{widetext}
which generalizes the well-known conservation equation   
$ \dot{\rho}+3H \left( P+\rho \right) =0$ valid in a FLRW 
universe, to which it reduces in the limit $m \rightarrow 0$. 
For the 
Schwarzschild interior solution there is no equation 
analogous to (\ref{1003}) because $H=0$ for the static Minkowski 
background and $\rho $ is static.

Eq.~(\ref{1002}) can be rewritten as 
\begin{equation} 
\partial_c P +u_c u^b \partial_b P +\left( P+\rho \right) u^b 
\nabla_b u_c=0 \;.
\end{equation}
By setting the index $c=1$ and computing the covariant 
derivative one 
obtains
\begin{equation} 
\frac{\partial P}{\partial r} +\left( P+\rho \right) 
\frac{m\bar{r}}{\bar{r}_0^3} \, \frac{
\left( 1 + \frac{m}{2\bar{r}_0} \right)}{
\left( 1+\frac{m\bar{r}^2}{2\bar{r}_0^3} \right) 
\left[ 1 - \frac{m}{\bar{r}_0} +
\frac{m\bar{r}^2}{\bar{r}_0^3} 
\left( 1 -\frac{m}{4\bar{r}_0} \right)    \right] 
}=0 \;.
\end{equation}
In the Newtonian limit $m/\bar{r}, m/\bar{r}_0 << 1$, $P<<\rho$, 
$r\simeq 
\bar{r}$, this equation reduces to 
\begin{equation} 
\frac{\partial P}{\partial r} +  
\frac{d\Phi_N }{dr} \, \rho = 0 \;,
\end{equation}
where $\rho= m\left( \frac{4\pi}{3} \, r_0^3 \right)^{-1}$ and 
$\Phi_N=\frac{m \bar{r}^2}{2\bar{r}_0^3}$ is the Newtonian 
potential. This expression for the density can also be obtained 
from eq.~(\ref{birba1}) by setting $a\equiv 1$ and using the 
curvature 
radius.  The first order correction to the equation of 
hydrostatic equilibrium for a spherically symmetric, uniform 
density star, is given by
\begin{equation} 
\frac{dP}{dr}+\frac{d\Phi_N}{dr}\, \rho \left\{ 
1-\frac{3}{2}\left[ \Phi_N(r)-\Phi_N(r_0) \right]\right\} =0 \;.
\end{equation}  

\subsection{The Schwarzschild-de Sitter black hole}

When the de Sitter space is chosen as the background, it is well 
known that the McVittie metric reduces to the Schwarzschild-de 
Sitter (K\"{o}ttler) metric, which can be put into the static 
form (\ref{15}). There is little doubt that in this case the 
central object 
described by this metric is a black hole. The Schwarzschild-de 
Sitter black hole has been the subject of much  
literature, mainly devoted to study the thermodynamical 
properties of dynamical horizons, which are interesting because 
of the simultaneous presence of a Schwarzschild and  a Rindler 
horizon. The inner horizon at $\bar{r}=m/2$ has area
\begin{equation} 
{\cal A}=\int\int d\theta d\varphi \sqrt{g_{\Sigma}} \;,
\end{equation}
where $g_{ab}\left.\right|_{\Sigma}$ is the restriction of the 
metric tensor to this 2-sphere and $g_{\Sigma}$ is its 
determinant. Eq.~(\ref{McVittie}) yields
\begin{equation} 
{\cal A}=16\pi a^2 m^2
\end{equation}
or, upon use of eq.~(\ref{36}), 
\begin{equation} \label{McVittiearea}
{\cal A}=16\pi \, m_H^2 \;.
\end{equation}
This area does not depend on time because $\dot{m}_H=0$ for all 
McVittie solutions as  a consequence of eq.~(\ref{35}). The 
Schwarzschild radial 
coordinate corresponding to the horizon is the curvature 
coordinate 
\begin{equation}  \label{McVittieradius}
r=\sqrt{ \frac{{\cal A} }{4\pi}}=2m_H 
\end{equation}
and is also time-independent.  Therefore, the horizon area of 
the  Schwarzschild-de Sitter black hole does not increase: 
the horizon does not expand, ``resisting'' the cosmic  
(accelerated) expansion.  Cosmic censorship (in its version for 
non-asymptotically flat spaces 
\cite{generalizedCosmicCensorship}) 
is not violated in this case; the 
central singularity remains forever and is always surrounded by 
a 
horizon.

It is not true that {\em  all} physical systems in {\em any} 
expanding universe 
participate in the  cosmic expansion; in a de Sitter universe 
this is  true only for 
weakly coupled systems. This result agrees with 
the ``all or nothing'' behaviour discovered by Price 
\cite{Price}. In the next section we argue that this phenomenon, 
however, is peculiar to the de Sitter background.

\section{Black holes in arbitrary FLRW backgrounds: old and 
new solutions}

Although the Schwarzschild-de  Sitter black hole does not 
participate in the cosmic expansion, 
in more general FLRW spacetimes black holes or other strongly  
gravitating objects  can expand under suitable conditions, as is 
shown below.

The first step toward this discussion consists of identifying 
exact inhomogeneous solutions of the Einstein equations that are  
suitable for describing strongly gravitating objects embedded in 
a FLRW universe that is not a de Sitter space. In this case the 
McVittie metric (\ref{McVittie}) can not be interpreted as  
describing black holes 
\cite{Sussman,Ferrarisetal,Nolan, McClureDyerCQG}, however, it 
does describe 
some kind of singular central  object. The computation of the 
area of the singular surface $\bar{r}=m/2$ and of its 
curvature radius proceeds exactly as in the 
Schwarzschild-de Sitter case leading 
again to the time-independent area (\ref{McVittiearea}) and 
radius (\ref{McVittieradius}). This applies also to the 
Reissner-Nordstrom generalization of McVittie's metric found by 
Gao and Zhang \cite{GaoZhang}. 
 The issue is whether this 
central 
object is a realistic one. It is tempting to 
interpret the singularity at $\bar{r}=m/2$ as an artificial 
one created by explicitly forbidding accretion of the cosmic 
fluid onto the central object, much like the axial singularity 
in cylindrically symmetric Bach-Weyl solutions \cite{BachWeyl}. 
In the latter,  
 two massive particle are in static 
equilibrium at a finite distance and an axial singularity is 
interpreted as a strut holding them apart \cite{Stephanietal}. 
One could think that the $\bar{r}=m/2$ surface in the McVittie 
metric similarly acts as a wall to keep out the cosmic 
fluid. This interpretation seems to be corroborated by the 
following observation:  the accretion rate for spherical 
accretion of a 
{\em test fluid} with energy  density $\rho$ and pressure $P $ 
onto a Schwarzschild black hole of mass $\mu$ is 
\cite{accretion}
\begin{equation} \label{testfluidaccretionrate}
\dot{\mu} =4\pi D \mu^2  \left( P_{\infty} +\rho_{\infty} 
\right) \;,
\end{equation}
where $\rho_{\infty} $ and $P_{\infty}$ are the energy density 
and pressure at spatial infinity, respectively, and $D$ is a 
constant corresponding to a first integral of motion 
\cite{accretion, BDE}. The 
accretion rate (\ref{testfluidaccretionrate}) vanishes for a 
fluid satisfying the quantum vacuum equation of state 
$P=-\rho=\frac{-\Lambda}{8\pi G}$. Extrapolating this result to 
a 
gravitating fluid, (spherical) accretion onto a 
Schwarzschild-de Sitter black hole is  automatically prevented, 
the hole's mass does not change, and the condition (\ref{35}) is 
satisfied naturally---it does no longer enforce  the presence of 
a  spherical wall to stop cosmic matter  from falling onto the 
black hole.

This interpretation of the 2-sphere singularity turns out to be 
at least partially 
erroneous. Below, some new 
exact solutions are presented in which the no-accretion 
condition 
is removed but  the pressure is still singular on the surface 
$\bar{r}=m/2$ unless the black hole is exactly comoving.

\subsection{The Sultana-Dyer solution}

Recently Sultana and Dyer \cite{SultanaDyer} found an exact 
solution of 
Petrov type~D describing a black hole embedded in a spatially 
flat FLRW universe with scale factor $ a(t)\propto t^{2/3}$ (in 
comoving time). The technique used to generate this solution 
consists of conformally transforming the Schwarzschild metric 
$g_{ab}^{(S)} \rightarrow \Omega^2 \, g_{ab}^{(S)}$ with the 
goal of changing the Schwarzschild global timelike Killing field 
$\xi^c$ into a  conformal Killing field for $\xi^c \nabla_c 
\Omega \neq 0$, generating the conformal Killing horizon (which 
differs from Hayward's future outer trapping horizons 
\cite{Hayward2} and from dynamical horizons 
\cite{dynhorizons}). The metric obtained by choosing 
$\Omega=a(t)=a_0 \, t^{2/3}$ --- the scale factor of a 
dust-filled 
$k=0$ FLRW universe --- is 
\begin{equation} \label{SultanaDyer}
ds^2=-\frac{  \left(1-\frac{m_0}{2\bar{r}} \right)^2}{
\left(1+\frac{m_0}{2\bar{r}} \right)^2} \, dt^2+
a^2(t) \left( 1+\frac{m_0}{2\bar{r}} \right)^4 \left( 
d\bar{r}^2 +\bar{r}^2 d\Omega^2 \right) \;;
\end{equation}
contrary to Ref.~\cite{SultanaDyer} we use isotropic radius 
$\bar{r}$ and comoving time $t$. 
This is formally the same as the McVittie solution 
(\ref{McVittie}) but with the important difference that the 
metric coefficient $m$ is now a constant $m_0$. To relate to our 
previous discussion, this implies that the 
Hawking quasi-local mass is $m_H(t)=m_0 \, a(t)$ and it {\em 
increases} with time. The condition (\ref{35}) is violated and 
an accretion flow of cosmic fluid onto the central object is 
present and is responsible for the increase in the gravitating 
mass.

The source for this metric is a combination of two 
non-interacting fluids: $T_{ab}= T_{ab}^{(I)}+T_{ab}^{(II)}$, 
where $ T_{ab}^{(I)}=\rho u_a\, u_b $ describes an ordinary 
(massive) dust and $ T_{ab}^{(II)}=\rho_n \, k_a\, k_b$ 
describes a null dust with density $\rho_n $ and $k^c k_c=0$ 
\cite{SultanaDyer}. The surface area of the conformal Killing 
horizon $\bar{r}=m_0/2$ is 
\begin{equation} 
{\cal A}(t)=\int\int d\theta d\varphi \, \sqrt{ 
g_{\Sigma}}=16\pi m_0^2 \, a^2(t)
\end{equation}
and its physical  (curvature coordinate) radius is simply
\begin{equation} 
r_{phys}(t)=\sqrt{ \frac{{\cal A}}{4\pi}}= 2 \, m_0 \, a(t) \;,
\end{equation}
which coincides with the familiar 
Schwarzschild radius $r=2m$ multiplied  
by the scale factor, as customary in FLRW cosmology. Thus, the 
physical radius of the horizon and the quasi-local mass are 
comoving with the cosmic background.

The Sultana-Dyer solution is non-singular at the surface 
$\bar{r}=m/2$ and can be interpreted as describing a black hole 
embedded in a two-fluid universe. We see that removing the 
McVittie no-accretion condition (\ref{35})  can indeed 
remove the singularity here, allowing  the black hole to become 
comoving with the rest of the universe. There are, however, 
problems with the Sultana-Dyer solution: the cosmological fluid 
becomes tachyonic (negative energy density) at late times near 
the horizon \cite{SultanaDyer}. Moreover, it is desirable to 
have cosmological matter described by a single fluid composed of 
particles following timelike geodesics, and to drop the 
restriction to the special choice of the scale factor $a\propto 
t^{2/3}$.

McClure and Dyer \cite{McClureDyerCQG} found another solution 
for a black hole-like object embedded in a radiation-dominated 
universe with a heat current, which satisfies the energy 
conditions everywhere and is perfectly comoving. However, the 
energy density and pressure are singular at $\bar{r}=m/2$. 
A similar solution in a dust-dominated universe has singular  
energy density \cite{McClureDyerCQG}.

It is not clear at this point whether 
solutions exist describing black holes in arbitrary FLRW 
backgrounds,  which are free of singularities at $\bar{r}=m/2$ 
and satisfy the energy conditions everywhere. This is the 
subject of the next subsection.

\subsection{New exact solutions}

We look for solutions described by a generalized  McVittie 
metric of the form (\ref{McVittie}), but without imposing the 
no-accretion restriction (\ref{35}),  and with arbitrary scale 
factor $a(t)$. The line element is written in the form
\begin{equation} 
ds^2= -\frac{B^2\left(t, \bar{r} \right)}{
A^2\left(t, \bar{r} \right)}\, dt^2 +a^2(t) A^4 \left(t, \bar{r} 
\right) \left( d\bar{r}^2+\bar{r}^2 d\Omega^2 \right) \;,
\end{equation}
where
\begin{equation}  
A \left(t, \bar{r} \right) = 1+\frac{m(t)}{2\bar{r}} \;, 
\;\;\;\;\;\;\;
B \left(t, \bar{r} \right) = 1-\frac{m(t)}{2\bar{r}} \;.
\end{equation}
The only non-vanishing components of the mixed Einstein tensor 
are 
\begin{eqnarray}
G_0^0 &=& -\, \frac{3A^2}{B^2}\left( 
\frac{\dot{a}}{a} +\frac{\dot{m}}{\bar{r}A} \right)^2 \;, 
\label{einst1} \\
&&\nonumber \\
G_0^1 &=&  \frac{2m}{ \bar{r}^2 a^2 A^5 B} \left( 
\frac{\dot{m}}{m} + \frac{\dot{a}}{a} \right) \;, \\
&&\label{einst2} \nonumber \\
G_1^1 &=& G_2^2 =G_3^3 \nonumber \\
&&\nonumber \\
&=&
-\, \frac{A^2}{B^2}\left\{ 
2  \frac{d}{dt} \left( 
\frac{\dot{a}}{a}+\frac{\dot{m}}{\bar{r}A} 
\right) 
+ \left( \frac{\dot{a}}{a}+\frac{\dot{m}}{\bar{r}A} \right) 
\right.\nonumber \\
&& \nonumber \\
&& \left. \cdot \left[
3 \left( \frac{\dot{a}}{a}+\frac{\dot{m}}{\bar{r}A} \right)
+ \frac{2\dot{m}}{\bar{r}AB} \right]\right\}  \;. \label{einst3}
\end{eqnarray}
It is convenient to introduce the quantity
\begin{equation} 
C\equiv  \frac{\dot{a}}{a}+\frac{\dot{m}}{\bar{r}A} =
\frac{\dot{m}_H}{m_H}-\frac{\dot{m}}{m} \frac{B}{A}\;,
\end{equation}
which reduces to $ \dot{m}_H/m_H $ for Sultana-Dyer-type 
solutions with $m=$constant. For any choice of the 
function $m(t)$ the quantity $C$ reduces to 
\begin{equation}  
C_{\Sigma}=
\frac{\dot{a}}{a}+\frac{\dot{m}}{m} =
\frac{\dot{m}_H}{m_H}  
\end{equation}
 on the surface $\bar{r}=m/2$. 
McVittie solutions have $C_{\Sigma}=0$ while 
Sultana-Dyer-type solutions have $C=C_{\Sigma}=H$ everywhere. 

The Ricci curvature  is 
\begin{equation} 
R=-g^{ab} G_{ab}= \frac{3A^2}{B^2}\left( 2\dot{C} +4C^2 +\frac{ 
2\dot{m}C}{\bar{r}AB} \right)
\end{equation}
and is singular on the surface $\bar{r}=m/2$ if the pressure is 
singular there. We now specialize the discussion to different 
forms of matter.

\subsubsection{Perfect fluid}

If we assume that matter is described by a single perfect fluid 
with  
stress energy tensor of the form
\begin{equation} 
T_{ab}=\left( P+\rho \right)u_a u_b +Pg_{ab} 
\end{equation}
and allow for a radial energy flow described by the fluid 
four-velocity $ u^c=\left( u^0, u, 0,0 \right) $, it is easy to 
see that the only possible solution is the Schwarzschild-de 
Sitter black hole already considered. In fact, the normalization 
$u^c u_c=-1 $ yields
\begin{equation} 
u^0=\frac{A}{B} \, \sqrt{ 1+a^2 A^4 u^2} 
\end{equation}
and the Einstein equations give, using 
eqs.~(\ref{einst1})-(\ref{einst3}) 
\begin{equation}  \label{delta0}
\dot{m}_H=-G B^2 au \left( P+\rho \right) {\cal A} \sqrt{ 
1+a^2A^4 u^2} \;,
\end{equation}
where  ${\cal A}=\int \int d\theta d\varphi \, 
\sqrt{g_{\Sigma}}=4\pi a^2 A^4 \bar{r}^2$ is the area of a 
spherical surface of isotropic  radius $\bar{r}$, 
\begin{eqnarray} 
&& 3\left( \frac{AC}{B} \right)^2=8\pi G \left[ \left( P+\rho 
\right)a^2A^4 u^2 +\rho \right] \;, \\
&&\nonumber \\
&& -\left( \frac{A}{B} \right)^2 \left( 2\dot{C}+3C^2 
+\frac{2\dot{m}C}{\bar{r}AB} \right)= \nonumber \\
&& =8\pi G \left[ \left( P+\rho \right)a^2A^4 u^2  +P \right] 
\;,    \label{delta1} \\  
&& \nonumber \\
&& -\left( \frac{A}{B} \right)^2 \left( 2\dot{C}+3C^2 
+ \frac{2\dot{m}C}{\bar{r}AB} \right)=8\pi G P  \;.
\label{delta2} 
\end{eqnarray}
By adding eqs.~(\ref{delta1}) and (\ref{delta2}) one obtains 
$P=-\rho$, i.e., only the de Sitter equation of state is 
allowed. Eq.~(\ref{delta0}) accordingly yields $\dot{m}_H=0$. 
With the exception of the non-accreting Schwarzschild-de Sitter 
black hole, 
a single perfect fluid can not source solutions representing 
spherically symmetric black holes embedded in a cosmological 
background. A mixture of two perfect fluids still constitutes a  
potential source, as exemplified by the Sultana-Dyer solution 
\cite{SultanaDyer}.

\subsubsection{Imperfect fluid and no radial mass flow}

The following solutions describe perfectly comoving black holes.  
We assume now that cosmological matter is described by the 
imperfect 
fluid stress-energy tensor 
\begin{equation} \label{imperfect}
T_{ab}=\left( P+\rho \right)u_a u_b +Pg_{ab}+ q_a u_b+q_b u_a 
\;,
\end{equation}
where the purely spatial vector $q^c$ describes a radial energy 
flow, 
\begin{equation} 
u^a=\left( \frac{A}{B}, 0,0,0 \right) \;, \;\;\;\;\;
q^b=\left( 0, q, 0,0 \right) \;, \;\;\;\;\; q^cu_c=0 \;,
\end{equation}
and $ u^c u_c=-1 $. The $\left( 0,1 \right)$ component of the 
Einstein equations 
$G_{ab}=8\pi G T_{ab}$ yields
\begin{equation} 
\frac{\dot{m}}{m}+\frac{\dot{a}}{a}= -\frac{4\pi G}{m}\, 
\bar{r}^2 a^2 A^4 B^2 q  \;.
\end{equation}
Since the Hawking mass is $m_H=m(t)a(t)$, it is
\begin{equation} 
\frac{\dot{m}_H}{m_H}=\frac{\dot{m}}{m}+\frac{\dot{a}}{a}
\end{equation}
and the area of a spherical surface of isotropic  radius 
$\bar{r}$ is ${\cal A}=\int \int d\theta d\varphi \, 
\sqrt{g_{\Sigma}}=4\pi a^2 A^4 \bar{r}^2$, yielding
the relation between energy flow, area ${\cal A}$, and 
accretion rate (Hawking mass added per unit time)
\begin{equation}  \label{accretionrate}
\dot{m}_H(t)=-G a B^2 {\cal A} q  \;.
\end{equation}
On a sphere of radius $\bar{r}>> m$ this 
can be written as (taking into account the fact that radial 
inflow corresponds to $q<0$)
\begin{equation}  \label{accretionrate2}
\dot{m}_H \simeq G a {\cal A} \left| q  \right|  \;.
\end{equation}
Hence, the quasi-local mass increases for inflow of matter. 

The $\left( 0,0 \right)$  and $\left( 1,1 \right)$ (or $\left( 
2, 2 \right)$ or $\left( 3,3\right)$)  components 
of the Einstein 
equations yield the energy density and pressure, respectively,
\begin{widetext}
\begin{eqnarray}
\rho \left( t, \bar{r}\right) &=&
\frac{1}{8\pi G} \, \frac{3A^2}{B^2} \left( 
\frac{\dot{a}}{a}+\frac{\dot{m}}{\bar{r}A} \right)^2 \;, 
\label{newdensity} \\
&&\nonumber \\
P \left( t, \bar{r}\right) &=&
\frac{-1}{8\pi G} \, \frac{A^2}{B^2} \left\{ 
2 \frac{d}{dt} \left( \frac{\dot{a}}{a}+\frac{\dot{m}}{\bar{r}A} 
\right) + 
\left( \frac{\dot{a}}{a}+\frac{\dot{m}}{\bar{r}A} 
\right)\left[ 3
\left( \frac{\dot{a}}{a}+\frac{\dot{m}}{\bar{r}A} 
\right)+\frac{2\dot{m}}{\bar{r}AB} \right]\right\}\;, \nonumber 
\\
&&  \label{newpressure}
\end{eqnarray}\end{widetext}
from which it is clear that the energy density is 
always non-negative. 
The expansion 
scalar is $3C$ and, in terms of this quantity,  
eq.~(\ref{newpressure}) becomes the generalized 
Raychaudhuri equation
\begin{equation} \label{generalizedRaychaudhuri}
\dot{C}=-\,\frac{3C^2}{2}-\frac{\dot{m}}{\bar{r}AB}\, C -4\pi G 
\,  \frac{B^2}{A^2} \, P \;.
\end{equation}
In the limit $m\rightarrow 0$ this reduces to the well known 
Raychaudhuri equation of FLRW cosmology 
$\dot{H}=-\frac{3H^2}{2}-4\pi G P$, for which the Hamiltonian 
constraint $H^2=8\pi G \rho/3$ then yields 
\begin{equation} \label{FLRWHdot}
\dot{H}=-4\pi G  \left(P+\rho \right) \;.
\end{equation}
Similarly, in the case $m\neq 0$, eq.~(\ref{newdensity}) yields
\begin{equation} 
\dot{C}=-4\pi G \, \frac{B^2}{A^2}\left( P+\rho \right) 
- \frac{\dot{m}C}{\bar{r}AB} \;,
\end{equation}
which generalizes eq.~(\ref{FLRWHdot}).

It can be noted that, due to the factor $B$ in the denominators, 
the energy density and pressure (\ref{newdensity}) and 
(\ref{newpressure}) appear to be singular on the surface 
$\bar{r}=m/2$ where $B$ vanishes. The situation is 
ameliorated by using the proper time $\tau$ defined  by 
$d\tau=\frac{B}{A} 
\, 
dt$ 
instead of the comoving time $t$ to absorb a factor $B$. This 
corresponds to using proper time instead of coordinate 
Schwarzschild time to offset an infinite redshift on the 
horizon of an ordinary Schwarzschild black hole, and yields 
\begin{equation}  
\frac{A}{B} \,  \left( \frac{\dot{a}}{a} 
+ \frac{\dot{m}}{\bar{r}A}\right) 
\longrightarrow 
\frac{a_{\tau}}{a}+\frac{m_{\tau} }{m}
 \;\;\;\;\;\;\;\;\;\; {\mbox as} \;\; \bar{r}\rightarrow 
m/2 \;,
\end{equation}
so that
\begin{equation} 
8\pi G \rho= 
3 \left( \frac{a_{\tau}}{a}+\frac{m_{\tau}}{\bar{r}A}  
\right)^2 
\end{equation}
and 
\begin{equation} 
8\pi G P = 
- 2 \left( \frac{a_{\tau}}{a}+\frac{ m_{\tau} 
}{\bar{r}A}  \right)_{\tau} -3 \left( 
\frac{a_{\tau}}{a}+\frac{m_{\tau} }{\bar{r}A}  
\right)^2   \;. \label{newpressure2}
\end{equation}
The pressure, the energy density, and the Ricci scalar 
$R=-\rho+3P$ appear to be finite on the surface $\bar{r}=m/2$. 
This is true for any form of the function $m(t)$ and contrasts 
with the singularities in the solutions of 
Ref.~\cite{McClureDyerCQG}.

Sultana-Dyer-type solutions with $\dot{m}=0$ 
have a conformal Killing horizon describing a cosmological 
black hole, as in the Sultana-Dyer 
\cite{SultanaDyer} solution. By design, this black hole is 
perfectly comoving: also in this case the cosmological expansion 
``wins'' over the local strong field of the black hole.
To keep $\dot{m}_H>0$ at $\bar{r}=m/2$, where $B=0$, 
one needs $q\rightarrow -\infty$ there. The Sultana-Dyer 
solution also suffers from a similar problem \cite{SultanaDyer}. 
This unphysical situation is due to the unrealistically  
simplified model of accretion.

\subsubsection{Imperfect fluid and radial mass flow}

We now consider an imperfect fluid with stress-energy tensor of 
the form (\ref{imperfect}) and both radial mass flow and energy 
current described by
\begin{equation} 
 u^a=\left( \frac{A}{B}\sqrt{1+a^2A^4u^2}, u, 0, 0  \right)\;, 
\;\;\;\;\;\;\;\;\;
q^c=\left( 0,q,0,0 \right) \;.
\end{equation}
By using the components (\ref{einst1})-(\ref{einst3}) of the 
Einstein tensor, the field equations become
\begin{widetext}
\begin{equation}  \label{deltadelta1}
\dot{m}_H=-GaB^2 {\cal A} \sqrt{1+a^2A^4u^2} \left[ \left( 
P+\rho 
\right)u+q \right] \;,
\end{equation}
\begin{equation} 
-3\left( \frac{AC}{B} \right)^2 =-8\pi G \left[ \left( P+\rho 
\right)a^2 A^4 u^2 +\rho \right] \;,
\end{equation}
\begin{equation} 
-\left( \frac{A}{B} \right)^2 \left( 2\dot{C}+3C^2 
+\frac{2\dot{m}C}{\bar{r}AB} \right)=8\pi G \left[ \left( P+\rho 
\right)a^2 A^4 u^2 +P+2a^2 A^4 qu \right] \;,
\end{equation}
\begin{equation} 
-\left( \frac{A}{B} \right)^2 \left( 2\dot{C}+3C^2 
+\frac{2\dot{m}C}{\bar{r}AB} \right)=8\pi G P \;.
\end{equation}
\end{widetext}
Adding the last two equations yields 
\begin{equation} \label{460}
q=-\left(P+\rho \right)\frac{u}{2} \;,
\end{equation}
i.e., to an ingoing radial flow of mass there corresponds an 
outgoing radial heat current if $P>-\rho$. By substituting 
eq.~(\ref{460}) into 
eq.~(\ref{deltadelta1}), one obtains the accretion rate
\begin{equation} 
\dot{m}_H=-\frac{G}{2} aB^2 \sqrt{1+a^2A^4 u^2} \left( P+\rho 
\right){\cal A}u \;,
\end{equation}
where $\left( P+\rho \right) {\cal A} u $ can be seen as the 
flux of gravitating energy through the surface of area ${\cal 
A}$. Since $u<0$, the mass $m_H$ increases if $P+\rho>0$, stays 
constant in a de Sitter background, and decreases if  
phantom energy with $P<-\rho$ is accreted. This lends support to 
the conclusions of Ref.~\cite{BDE} on the fate of  a black hole 
in a phantom-dominated universe.

Moreover, the energy density is given by
\begin{equation} 
8\pi G \rho= \frac{A^2}{B^2}\left[ 3C^2 +\left( 
\dot{C}+\frac{\dot{m}C}{\bar{r}AB} \right) 
\frac{2 a^2A^4u^2}{1+a^2A^4 u^2} \right] \;.
\end{equation}
For Sultana-Dyer-type solutions with $m=m_0=$constant the 
energy density reduces to
\begin{equation} 
8\pi G \rho= \frac{A^2}{B^2} \left[ 3H^2 +\frac{2\dot{H}a^2 A^4 
u^2}{1+a^2 A^4 u^2} \right]
\end{equation}
and is positive-definite in a superaccelerating universe with 
$\dot{H}>0$, which is necessarily phantom-dominated 
\cite{superquintessence}. Moreover, 
one can solve for the velocity of the fluid obtaining
\begin{equation} 
u=-\left\{ \frac{  \sqrt{ 1+\frac{4m_0^2H^2a^2A^4}{G^2B^4{\cal 
A}^2\left(P+\rho \right)^2}}-1}{2a^2A^4} \right\}^{1/2} \;.
\end{equation}
The motion of the fluid becomes superluminal as 
$\bar{r}\rightarrow m_0/2$, where $B\rightarrow 0$. In a  
realistic model the flow becomes supersonic at a certain 
distance from this surface. This fact can only be taken into 
account in a more realistic model of accretion, which will be 
studied elsewhere.

\section{Discussion and conclusions}

It is well known that the effect of the cosmological expansion 
on weakly gravitating Newtonian systems of small size $r$ 
(i.e., $rH_0<<1$) is completely negligible for practical 
purposes, even though these systems do participate in the 
expansion. In larger systems such as voids, filaments, and large 
scale structures, the 
cosmic expansion plays a significant role 
\cite{NoerdlingerPetrosian,  SatoMaeda, Bushaetal, 
BalagueraNowakowski}. But the size of the local system is not 
the only relevant factor. The study by Price \cite{Price} is the 
first to focus on the strength with which the local system is 
bound. The ``all or nothing'' behaviour discovered constitutes 
an important step in the understanding of the process. However, 
Price's discussion is limited to a de Sitter background which is 
too special to draw general conclusions. Moreover, the classical 
non-relativistic ``atom'' considered can not describe 
arbitrarily strong binding of the local system because, when the 
energy density and stresses involved become very large, they  
distort spacetime causing the  metric  to substantially  
deviate from a cosmological one. It is preferable to study exact 
solutions describing a local object with a  strong 
local gravitational field --- e.g., a black hole --- embedded in 
an expanding universe. 

Independent motivation for our study arises from the recent 
realization that if the current acceleration of the universe is 
dominated by phantom dark energy with $P<-\rho$, then the 
universe may be running into a Big Rip at a finite time in the 
future \cite{BigRip}. Recent literature  has focused on the way 
this catastrophic accelerated expansion of the universe comes to 
dominate the local dynamics of bound systems (clusters, galaxies, 
 stars, {\em etc.}) and tears these systems apart. In this 
context, it is interesting to pose the question of whether the 
Big Rip can destroy a black hole horizon and expose the central 
singularity, thus violating cosmic censorship 
\cite{generalizedCosmicCensorship}. A partial answer comes from 
Ref.~\cite{BDE}: these authors analyse spherical accretion of  
a phantom test fluid onto a Schwarzschild black hole and, 
extrapolating the results to a gravitating fluid, reach the 
conclusion that the horizon disappears before the Big Rip 
together with the central singularity, without violating cosmic 
censorship. This result is quite plausible, however it is 
desirable to have a study of the accretion process for  a 
gravitating fluid, which brings us again to the realm of exact 
solutions.

The fully relativistic systems considered in the present paper 
provide at least a partial answer to the questions above. The 
McVittie solution (\ref{McVittie}) and (\ref{35}) 
\cite{McVittie} is 
accretion-free, describes a 
general FLRW background universe, and does not expand. However, 
it has a mild singularity on the surface $\bar{r}=m/2$ (the 
putative horizon) \cite{Sussman,Ferrarisetal, Nolan} and 
therefore it does not describe a black hole. There is an 
important exception, the Schwarzschild-de Sitter black hole (a 
special case of the McVittie metric) which does not suffer from 
this singularity and describes a true black hole horizon which 
resists the cosmic expansion. This feature is consistent with 
Price's study of the classical atom in a de Sitter background 
\cite{Price} and with our phase space picture of Sec.~2.3.

The overall picture emerging is that the consideration of a de 
Sitter background is rather misleading, even though it 
considerably simplifies the calculations: the ``all or nothing'' 
behaviour is not generic of FLRW space but is limited to a de 
Sitter background. The next system considered, the Nolan 
interior solution \cite{NIS}, does not suffer from the 
singularity problem of McVittie's metric  because the surface 
$\bar{r}=m/2$ is covered by the matter composing the star. This 
solution, which can be thought of as providing a source for 
McVittie's metric to which it is matched, does not accrete 
either and can be embedded in a general FLRW background: it is 
comoving. Removing the limiting assumption of a de Sitter 
background  allows the strongly gravitating central object to 
expand. 

Another exact solution describing a black hole embedded in a 
non-de Sitter universe is the one of  Sultana and Dyer 
\cite{SultanaDyer} in which, contrary to the McVittie metric, 
accretion onto the black hole does occur and the conformal 
Killing 
horizon (the black hole horizon) is, by design, comoving. The 
peculiarity of the de Sitter cosmos is thus further put 
into evidence.  The Sultana-Dyer solution, however, suffers from 
the following limitations: i) it is restricted to the special 
form of the scale factor $a(t)=a_0 t^{2/3}$ in comoving time; 
ii) the matter source is not  a simple fluid but a 
mixture of two non-interacting fluids, one of which is a null 
dust; and iii) the energy density becomes negative at late times 
near the horizon. It is well known that, in general, matching 
black hole and cosmological metric produces stress-energy 
tensors that violate the energy conditions in some regions of 
the spacetime manifold \cite{McClureDyerCQG}.

To overcome these difficulties we have 
studied new alternative solutions of the Einstein equations 
which generalize the McVittie metric by allowing radial 
accretion onto the central object. We have presented new 
solutions for which the surface $\bar{r}=m/2$ is 
non-singular and is perfectly comoving. These new solutions do 
not always suffer from the limitations i)-iii) above---in 
particular,  the energy density is everywhere positive at all 
times for one  of the solutions, but the accretion flow 
generally becomes 
superluminal, an artifact of the simplified ``rigid'' model of 
accretion used, in which matter everywhere in the universe 
moves toward the central black hole (albeit its radial speed 
becomes zero far away from it). 
Moreover, these solutions support the result of Ref.~\cite{BDE} 
that a black hole embedded in a phantom-dominated universe 
disappears, respecting cosmic censorship.

It  appears, therefore, that the strong 
external cosmological 
gravitational field can distort a black hole horizon, or anyway 
stretch an object dominated by a strong local field.  
de Sitter-like expansion does not, but this should be seen as an 
exception to the rule due to the scale-invariant nature of the 
exponential scale factor (in fact, the de Sitter metric can be 
put into static form).  {\em A posteriori}, the fact that a  
black hole horizon 
expands with the cosmic substratum should not be regarded as 
surprising: many exact solutions are known in which a black hole 
horizon is distorted by its surroundings, due to an external 
gravitational or acceleration field 
\cite{KernsWild,FarooshZimmerman,BiniCherubiniMashhoon, 
Poisson, Anninosetal, Bretonetal, PatelTrivedi, Death, 
Doroshkevichetal}, an 
electric \cite{Ernst} or 
magnetic  \cite{ErnstWild, 
WildKerns,KulkarniDadhich,Krori} field, or 
combinations of them  \cite{Bicak}. A substantial amount of 
literature has been devoted to 
horizon deformation  (\cite{FrolovNovikov} and 
references therein).

Our results do not constitute the last word on the issue of 
cosmological expansion and local systems.  Future endeavours 
include the search for more general and more realistic exact 
solutions of the  Einstein equations describing accreting black 
holes in arbitrary FLRW or Bianchi backgrounds.

\begin{acknowledgments}
V.F. thanks Hideki Maeda for  a helpful discussion. This 
work  was supported by the Natural Sciences and Engineering  
Research  Council of Canada ({\em NSERC}).
\end{acknowledgments}


\begin{thebibliography}{99}

\bibitem{McVittie} G.C. McVittie, {\em Mon. Not. R. Astr. Soc.} 
{\bf 93}, 325 (1933).

\bibitem{EinsteinStraus} A. Einstein and E.G. Straus, {\em Rev. 
Mod. Phys.} {\bf 17}, 120 (1945); {\bf 18}, 148 (1946).

\bibitem{Bonnoratom} W.B. Bonnor, {\em Class. Quant. Grav.} {\bf 
16}, 1313 (1999).

\bibitem{Krasinski} A. Krasinski, {\em Inhomogeneous Cosmological 
Models} (CUP, Cambridge, 1997).

\bibitem{non-extension} J.M.M. Senovilla and R. Vera, {\em Phys. 
Rev. Lett.} {\bf 78}, 2284 (1997); M. Mars, {\em Phys. Rev. D} 
{\bf 57}, 3389 (1998).

\bibitem{NolanPRD} B.C. Nolan, {\em Phys. Rev. D} {\bf 58}, 
064006 (1998).

\bibitem{variousworks} 
J. Pachner, {\em Phys. Rev.} {\bf 132}, 1837 (1963); {\em Phys. 
Rev. B} {\bf 137}, 1379 (1965); 
W.M. Irvine, {\em Ann. Phys.} {\bf 32}, 322 (1965);
R.H. Dicke and P.J.E. Peebles, {\em Phys. Rev. Lett.} {\bf 12}, 
435 (1964);
C. Callan, R.H. Dicke, and 
P.J.E. Peebles, {\em Am. J. Phys.} {\bf 33}, 105 (1965); 
P. D'Eath, {\em Phys. Rev. D} {\bf 11}, 1387 (1975);
R.P.A. Newman and G.C. McVittie, {\em Gen. Rel. Grav.} {\bf 14}, 
591 (1982);
R. Gautreau, {\em Phys. Rev. D} {\bf 29}, 198 (1984);
P.A. Hogan, {\em Astrophys. J.} {\bf 360}, 315 (1990); 
J.L. Anderson, {\em Phys. Rev. Lett.} {\bf 75}, 3602 (1995);
W.B. Bonnor, {\em Mon. Not. R. Astr. Soc.} {\bf 282}, 1467 
(1996);
A. Feinstein, J. Ibanez, and R. Lazkoz, {\em Astrophys. J.} {\bf 
495}, 131 (1998);
K.R. Nayak, M.A.H. MacCallum, and C.V. Vishveshwara, {\em 
Phys. Rev. D} {\bf 63}, 024020 (2000);
V. Guruprasad, gr-qc/0005090; gr-qc/0005014;
G.A. Baker Jr., astro-ph/0003152;
A. Dominguez and J. Gaite, {\em Europhys. Lett.} {\bf 55}, 458 
(2001);
T.M. Davis and C.H. Lineweaver, {\em AIPC} {\bf 555}, 348 
(2001);
G.F.R. Ellis, gr-qc/0102017;
C. Stornaiolo, {\em Gen. Rel. Grav.} {\bf 34}, 2089 (2002);
T.M. Davis, C.H. Lineweaver, and J.K. Webb, {\em Am. J. Phys.} 
{\bf 71}, 358 (2003);
L. Lindegren and D. Dravins, {\em Astron. Astrophys.} {\bf 401}, 
1185 (2003);
C.J. Gao, {\em Class. Quant. Grav.} {\bf 21}, 4805 (2004);
T.M. Davis and C.H. Lineweaver, {\em Publ. Astr. Soc. Pac.} {\bf 
21}, 97 (2004);
T.M. Davis, PhD thesis, Univ. of New South 
Wales (astro-ph/0402278);
D.P. Sheehan and V.G. Kriss, astro-ph/0411299;
W.J. Clavering, astro-ph/0511709;
Z.-H. Li and A. Wang, astro-ph/0607554;
O. Gron and O. Elgaroy, astro-ph/0603162;
L.A. Barnes, M.J. Francis, J. B. James, and G.F. Lewis, 
{\em Mon. Not. R. Astr. Soc.} {\bf 373}, 382 (2006);
R. Lieu and D.A. Gregory, astro-ph/0605611;
P.K.F. Kuhfittig, gr-qc/0608120;
G.S. Adkins, J. McDonnell, and R.N. Fell, {\em Phys. Rev. D} 
{\bf 
75}, 064011 (2007);
D.L. Wiltshire, gr-qc/0702082;
M. Sereno and P. Jetzer, {\em Phys. Rev. D} {\bf 75}, 064031 
(2007).

\bibitem{NoerdlingerPetrosian} P.D. Noerdlinger and V. 
Petrosian, {\em Astrophys. J.} {\bf 168}, 1 (1971).

\bibitem{SatoMaeda} H. Sato and K. Maeda, {\em Prog. Theor. 
Phys.} {\bf 70}, 119 (1983).

\bibitem{Sussman} R. Sussman, {\em Gen. Rel. Grav.} {\bf 17}, 
251 (1985).

\bibitem{Death} P.D. D'Eath, {\em Phys. Rev. D} {\bf 11}, 1387 
(1975).

\bibitem{Ferrarisetal} M. Ferraris, M. Francaviglia, and A. 
Spallicci, {\em Nuovo Cimento} {\bf 111B}, 1031 (1996).

\bibitem{CFV} F.I. Cooperstock, V. Faraoni, and D.N. Vollick, 
{\em Astrophys. J.} {\bf 503}, 61 (1998).

\bibitem{Bombelli} B. Bolen, L. Bombelli, and R. Puzio, {\em 
Class. Quant. Grav.} {\bf 18}, 1173 (2001).

\bibitem{Nolan} B.C. Nolan, {\em Class. Quant. Grav.} {\bf 16}, 
1227 (1999).

\bibitem{Nolan2} B.C. Nolan, {\em Class. Quant. Grav.} {\bf 16}, 
3183 (1999).

\bibitem{Bushaetal} M.T. Busha, F.C. Adams, R.H. 
Wechsler, and 
A.E. Evrard, {\em Astrophys. J.} {\bf 596}, 713 (2003).

\bibitem{GaoZhang} C.J. Gao and S.N. Zhang, {\em Phys. Lett. B} 
{\bf 595}, 28 (2004).

\bibitem{SultanaDyer} J. Sultana and C.C. Dyer, {\em Gen. Rel. 
Grav.} {\bf 37}, 1349 (2005).

\bibitem{Pioneer}
M. Mizony and M. Lachi\`{e}ze-Rey, gr-qc/0412084;
D. Izzo and A. Rathke, astro-ph/0504634;
F.J. Oliveira, gr-qc/0610029;
M. Lachi\`{e}ze-Rey, gr-qc/0701021;
J. Rosales and J. Sanchez-Gomez, gr-qc/99810085;
M. Carrera and D. Giulini, gr-qc/0602098; gr-qc/0605078; 
C. Lammerzhal and O. Preuss, gr-qc/0604052;
S.G. Turyshev  and J.G. Williams, gr-qc/0611095;
H.-J. Fahr and M. Siewert, gr-qc/0610034;
J.G. Williams, S.G. Turyshev, and D.H. Boggs, {\em Phys. Rev. 
Lett.} {\bf 98}, 059002 (2007);
Y.V. Dumin, gr-qc/0610035.

\bibitem{NesserisPerivolaropoulos} S. Nesseris and L. 
Perivolaropoulos, {\em Phys. Rev. D} {\bf 70}, 123529 (2004).

\bibitem{Price} R.H. Price, gr-qc/0508052.

\bibitem{BalagueraNowakowski} A. Balaguera-Antolinez and M. 
Nowakowski, gr-qc/0704.1871.

\bibitem{Mashhoonetal} B. Mashhoon, N. Mobed, and D. Singh, 
arXiv:0705.1312.

\bibitem{reviews}  
J.M. Senovilla, M. Mars, and R. Vera, {\em Phys. World}, July 
1999, 20 (1999);
W.B. Bonnor, {\em Gen. Rel. Grav.} {\bf 38}, 1005 (2000).

\bibitem{primordialBHs} Y.B. Zel'dovich and I.D. Novikov, {\em 
Sov. Astr. A.J.} {\bf 10}, 602 (1967); S.W. Hawking, {\em Phys. 
Rev. Lett.} {\bf 26}, 1344 (1971); {\em Mon. Not. R. Astr. 
Soc.} {\bf 152}, 75 (1971);
B.J. Carr and S.W.  Hawking, {\em Mon. Not. R. Astr. Soc.} {\bf 
168}, 399 (1974); N. Sakai and J.D> Barrow, {\em Class. Quant. 
Grav.} {\bf 18}, 4717 (2001).

\bibitem{HaradaCarrMaeda} T. Harada and B.J. Carr, {\em Phys. 
Rev. D} {\bf 71}, 104009 (2005); {\bf 71}, 104010 (20050; {\bf 
72}, 044021 (2005); T. Harada, H. Maeda, and B.J. Carr, {\em 
Phys. Rev. D}  {\bf 74}, 024024 (2006); B.J. Carr, 
astro-ph/0511743.

\bibitem{SaidaHaradaMaeda}  H. Saida, T. Harada, and H. Maeda, 
arXiv:0705.4012. 

\bibitem{SN}  A.G. Riess {\em et al.} 1998, {\em Astron. J.} 
{\bf 116}, 1009; 1999, {\em Astron. J.} {\bf 118}, 2668;
2001, {\em Astrophys. J.} {\bf 560}, 49;
2004, {\em Astrophys. J.} {\bf 607}, 665;
S. Perlmutter {\em et al.}, {\em Nature} {\bf 391}, 51 
(1998); {\em Astrophys. J.} {\bf 517}, 565 (1999); 
J.L. Tonry {\em et al.}, {\em Astrophys. J.} {\bf 594}, 
1 (2003);
R. Knop {\em et al.}, {\em Astrophys. J.} {\bf 598}, 102 
(2003);
B. Barris {\em et al.}, {\em Astrophys. J.} {\bf 602}, 
571 (2004);
A.G. Riess {\em et al.}, astro-ph/0611572

\bibitem{BigRip} R.R. Caldwell, M. Kamionkowski, and N.N. 
Weinberg, {\em Phys. Rev. Lett.} {\bf 91}, 071301 (2003).

\bibitem{generalizedCosmicCensorship} B.J. Carr and S.W. 
Hawking, {\em Mon. Not. R. Astr. Soc.} {\bf 168}, 399 (1974); 
G.W. Gibbons and S.W. Hawking, {\em Phys. Rev. D} {\bf 15}, 2738 
(1977); R. Penrose, {\em Gravitational Collapse}, in {\em IAU 
Symposium n.~64} (Reidel, Dordrecht, 1974); {\em Singularities 
of  Space-Time}, in {\em Theoretical Principles in Astrophysics 
and  Relativity}, N.R. Lebovitz, W.H. Reidel, and P.O. 
Vandecoord  eds. (University of Chicago Press, 1978); A. Krolak, 
{\em Gen.  Rel. Grav.} {\bf 16}, 121 (1984).

\bibitem{BDE} E. Babichev, V. Dokuchaev,  and Yu.  Eroshenko,  
{\em Phys. Rev. Lett.}  {\bf 93}, 021102 (2004).

\bibitem{NIS} B.C. Nolan, {\em J. Math. Phys.} {\bf 34}, 1 
(1993).
 
\bibitem{Wald} R.M. Wald, {\em General Relativity} (Chicago 
Univ. Press, Chicago, 1984).

\bibitem{footnote1} The  authors of Ref.~\cite{CFV} do 
not include the centrifugal term $L^2/r^3$ which is, however, 
straightforward to reintroduce. 

\bibitem{phantom} 
S. Capozziello, S. Nojiri, and S.D. Odintsov, {\em 
Phys. Lett. B} {\bf 632}, 597 (2006);
S. Nojiri and S.D.  Odintsov , hep-th/0506212; 
{\em Phys. Rev. D} {\bf 72}, 023003 (2005);
V. Faraoni, {\em Class. Quantum Grav.} {\bf 22}, 3235 
(2005); 
W. Fang {\em et al.}, {\em Int. J. Mod. Phys. D} {\bf 
15}, 199 (2006); 
M.G. Brown, K. Freese, and W.H. Kinney, astro-ph/0405353;
E. Elizalde, S. Nojiri, and S.D. Odintsov, {\em Phys. 
Rev. D} {\bf 70},  043539 (2004); 
{\em Phys. Lett. B} {\bf 574}, 1 (2003); 
{\em Phys. Rev. D} {\bf 70}, 043539 (2004);  
J.-G. Hao and  X.-Z. Li, {\em Phys. Lett. B} {\bf 606}, 
7 (2005);
{\em Phys. Rev. D} {\bf 68}, 043501 (2003);
X.-Z. Li and J.-G. Hao, 
{\em Phys. Rev. D} {\bf 69}, 107303 (2004);
J.M. Aguirregabiria, L.P.  Chimento, and R. Lazkoz, {\em 
Phys. Rev. D} {\bf 70},  023509 (2004);
Y.-S. Piao  and Y.-Z. Zhang, {\em Phys. Rev. D} {\bf 
70}, 063513 (2004);
H.Q. Lu, {\em Int. J. Mod. Phys. D} {\bf 14}, 355 (2005);
V.B. Johri, {\em Phys. Rev. D} {\bf 70}, 041303(R) (2004);
H. Stefanci\'{c}, {\em Phys. Lett. B} {\bf 586}, 5 (2004);
D.J. Liu and X.-Z. Li, {\em Phys. Rev. D} {\bf 68},  
067301 (2003);
X.-Z. Li and J.G. Hao, {\em Phys. Rev. D} {\bf 69},  
107303 (2004);
M.P. Dabrowski, T. Stachowiak, and M. Szydlowski, {\em 
Phys. Rev. D} {\bf 68},  103519 (2003);
Z.K. Guo, Y.S. Piao, and  Y.Z. Zhang, {\em Phys. Lett. 
B} {\bf 594}, 247 (2004);
J.M. Cline, S. Jeon, and G.D. Moore,  
  {\em Phys. Rev. D} {\bf 70}, 043543 (2004); 
S. Nojiri  and S.D.  Odintsov, {\em Phys. Lett. B} {\bf 
562}, 147 (2003);
L. Mersini, M. Bastero-Gil, and P. Kanti, {\em Phys. Rev. 
D} {\bf 64} 043508 (2001); 
M. Bastero-Gil, P.H. Frampton, and L. Mersini, 
  {\em Phys. Rev. D} 65 106002 (2002); 
P.H. Frampton, {\em Phys. Lett. B} {\bf 555}, 139 (2003).

\bibitem{KernsWild} R.M. Kerns and W.J. Wild, {\em Gen. Rel. 
Grav.} {\bf 14}, 1 (1982).

\bibitem{FarooshZimmerman} H. Farhoosh and R.L. Zimmerman, {\em 
Phys. Rev. D} {\bf 21}, 317 (1980).

\bibitem{BiniCherubiniMashhoon} D. Bini, C. Cherubini, 
and B. Mashhoon, gr-qc/0410098.

\bibitem{DhurandharDadhich} S.V. Dhurandhar and N. Dadhich, 
{\em Phys. Rev. D} {\bf 29}, 2712 (1984); {\bf 30}, 1625 (1984).

\bibitem{Poisson} E. Poisson, gr-qc/0501032;

\bibitem{Anninosetal} P. Anninos, D. Hobill, E. Seidel, L. 
Smarr, and W.-M. Suen, {\em Phys. Rev. Lett.} {\bf 71}, 2851 
(1993).

\bibitem{Bretonetal} N. Breton, A.A. Garcia, V.S. Manko, and 
T.E. Denisova, {\em Phys. Rev. D} {\bf 57}, 3382 (1998).

\bibitem{PatelTrivedi} L.K. Patel and H.B. Trivedi, {\em J. 
Astrophys. Astr.} {\bf 3}, 63 (1982).

\bibitem{Doroshkevichetal} A.G. Doroshkevich, Ya. B. Zel'dovich, 
and I.D. Novikov, {\em Zh. Eksp. Teor. Fiz.} {\bf 49}, 170 
(1965).

\bibitem{Ernst} F.J. Ernst, {\em J. Math. Phys.} {\bf 15}, 515 
(1976); {\bf 17}, 54 (1976).

\bibitem{ErnstWild} F.J. Ernst and W.J. Wild, {\em J. Math. 
Phys.} {\bf 17}, 182 (1976).

\bibitem{WildKerns} W.J. Wild and R.M. Kerns, {\em Phys. Rev. D} 
{\bf 21}, 332 (1980).

\bibitem{KulkarniDadhich} R. Kulkarni and N. Dadhich, {\em 
Phys. Rev. D} {\bf 33}, 2780 (1986).
 
\bibitem{Krori} K.D. Krori, {\em J. Math. Phys.} {\bf 25}, 607 
(1984).

\bibitem{Bicak} J. Bicak, {\em Proc. R. Soc. Lond. A} {\bf 371}, 
429 (1980). 

\bibitem{FrolovNovikov} V.P. Frolov and I.D. Novikov, {\em Black 
Hole Physics} (Kluwer Academic, Dordrecht, 1998).

\bibitem{Hawking} S.W. Hawking, {\em J. Math. Phys.} {\bf 9}, 
598 (1968).

\bibitem{Hayward} S.A. Hayward, {\em Phys. Rev. D} {\bf 49}, 
831 (1994).

\bibitem{McClureDyerCQG} M.L. McClure and C.C. Dyer, {\em Class. 
Quant. Grav.} {\bf 23}, 1971 (2006).

\bibitem{McClureDyerGRG} M.L. McClure and C.C. Dyer, {\em Gen. 
Rel. Grav.} {\bf 38}, 1347 (2006).

\bibitem{Thakurta} S.N.G. Thakurta, {\em UIndian J. Phys. B} 
{\bf 55}, 304 (1981).

\bibitem{Vaidya} P.C. Vaidya, {\em Pramana} {\bf 8}, 512 (1977).

\bibitem{Kustaanheimo} P. Kustaanheimo, {\em Comment. Phys. 
Math. Helsingf.} {\bf 13}, 8 (1947).

\bibitem{BachWeyl} R. Bach and H. Weyl, {\em Math. Z.} {\bf 13}, 
134 (1922).

\bibitem{Stephanietal} H. Stephani, D. Kramer, M. MacCallum, C. 
Hoenselaers, and E.  Herlt, {\em Exact Solutions of Einstein's 
Field Equations}, 2nd edition (CUP, Cambridge, 2003).

\bibitem{accretion} F.C. Michel, {\em Astrophys. Sp. Sci.} {\bf 
15}, 153 (1972); 
B.J. Carr and S.W. Hawking, {\em Mon. Not. R. Astr. Soc.} {\bf 
168}, 399 (1974); 
M.C. Begelman, {\em Astron. Astrophys.} {\bf 70}, 583 (1978); 
D. Ray, {\em Astron. Astrophys.} {\bf 82}, 368 (1980);
K.S. Thorne, R.A. Flammang, and A.N. Zytkow, {\em Mon. Not. R. 
Astr. Soc.} {\bf 194}, 475; E. Bettwieser and W. Glatzel, {\em 
Astron. Astrophys.} {\bf 94}, 306 (1981); 
K.M. Chang, {\em Astron. Astrophys.} {\bf 142}, 212 (191985);
U.S. Pandey, {\em Astrophys. Sp. Sci.} {\bf 136}, 195 (1987);
L.I. Petrich, S.L. Shapiro, and S.A. Teukolsky, {\em Phys. Rev. 
Lett.} {\bf 60}, 1781 (1988).

\bibitem{Hayward2} S.A. Hayward, {\em Phys. Rev. D} {\bf 49}, 
6467 (1994).

\bibitem{dynhorizons} A. Ashtekar and B. Krishnan, {\em Phys. 
Rev. Lett.} {\bf 89}, 261101 (2002).

\bibitem{superquintessence} V. Faraoni, {\em Int. J. Mod. Phys. 
D} {\bf 11}, 471 (2002).

\end{thebibliography}

\end{document}